\documentclass[journal,final,twocolumn,10pt,twoside]{IEEEtranTCOM}\ifCLASSINFOpdf
\else
\fi

\IEEEoverridecommandlockouts

\usepackage{amsmath}
\usepackage{subfig}

\usepackage{graphicx}
\usepackage{amsfonts}
\usepackage{bm}
\usepackage{wrapfig}
\usepackage{makeidx}
\usepackage{moreverb}
\usepackage[lined,boxed,commentsnumbered]{algorithm2e}
\usepackage{multirow}
\usepackage{array}
\usepackage{MnSymbol}
\usepackage{bbding}
\newtheorem{Alg}{Algorithm}
\usepackage{hyperref}
\usepackage{color}
\usepackage[usenames,dvipsnames]{xcolor}

\DeclareGraphicsExtensions{.jpg,.png,.tif,.eps}

\begin{document}

\title{Sparse Attack Construction and State Estimation in \MakeLowercase{t}he Smart Grid: Centralized and Distributed Models}

\author{
\authorblockN{Mete Ozay$^{\dag *}$, I\~naki Esnaola$^\dag$, Fatos T. Yarman Vural$^*$, Sanjeev R. Kulkarni$^\dag$ and H. Vincent Poor$^\dag$} \\
\authorblockA{$^\dag$ Department of Electrical Engineering, Princeton University, Princeton, NJ 08544, USA \\ \{mozay, jesnaola, kulkarni, poor\}@princeton.edu} \\
\authorblockA{$^*$ Department of Computer Engineering, Middle East Technical University, Ankara, Turkey {vural@ceng.metu.edu.tr}}
\thanks{Manuscript received September 21, 2012; revised March 15, 2013.}
\thanks{This research was supported in part by the Center for Science of Information (CSoI), an NSF Science and Technology Center, under Grant CCF-0939370, by the U. S. Air Force office of Scientific Research under MURI Grant FA9550-09-1-0643, and by the U. S. Army Research Office under MURI Grant W911NF-11-1-0036 and Grant W911NF-07-1-0185.}}



%


\maketitle

\begin{abstract}

New methods that exploit sparse structures arising in smart grid networks are proposed for the state estimation problem when data injection attacks are present. First, construction strategies for unobservable sparse data injection attacks on power grids are proposed for an attacker with access to all network information and nodes. Specifically, novel formulations for the optimization problem that provide a flexible design of the trade-off between performance and false alarm are proposed. In addition, the centralized case is extended to a distributed framework for both the estimation and attack problems. Different distributed scenarios are proposed depending on assumptions that lead to the spreading of the resources, network nodes and players. Consequently, for each of the presented frameworks  a corresponding optimization problem is introduced jointly with an algorithm to solve it. The validity of the presented procedures in real settings is studied through extensive simulations in the IEEE test systems. 

\end{abstract}




%
\IEEEpeerreviewmaketitle

\section{Introduction}

Power networks are complex systems consisting of generators and loads that are connected by transmission and distribution lines \cite{s1}. These systems can be modeled by complex networks, in which the generators and loads are represented by physically distributed nodes and power lines are represented by edges that connect the nodes. Because of the geographic and physical distribution of the nodes and the power transmission constraints \cite{lalitha}, various structural properties of complex networks are observed in power networks \cite{cn}. For instance, the distribution of electrical distances of Eastern, Western and Texas interconnects in the North American power network obeys a power-law distribution, which leads to scale free and hierarchical network structures \cite{cn}. 

The aforementioned structural properties of power networks constrain the way in which both attack and defense schemes are designed for the smart grid. Several attack vector construction and detection methods have been introduced using either centralized \cite{s1,kim,kosut,book} or distributed \cite{dist, OEYKP_smc_12_2} models. Data sparsity properties have been analyzed for constructing unobservable sparse attack vectors by Liu et al. \cite{s1}. Kosut et al. \cite{kosut} have introduced the relationship between attack detectability and network observability using a graph-theoretic model. Xie et al. \cite{fully} proposed a distributed wide-area state vector estimation algorithm which is also employed for bad data detection \cite{dist}. However, they do not exploit the sparsity and instead they define the state estimation problem as a weighted least squares (WLS) problem. Pasqualetti et al. \cite{dist2} solved a similar WLS problem using a measurement distributed decomposition method for distributed state estimation and attack detection. Yang et al. \cite{hierer} have proposed a hierarchical architecture to construct sparse attack vectors using combinatorial search methods. Vukovic et al. \cite{mitig} have analyzed various mitigation schemes of data integrity attacks for state estimation. Recent advances for attack vector construction and state vector estimation methods in power systems have been reviewed in \cite{review1} and \cite{review2}.

The centralized attack schemes proposed in this paper follow the undetectability criteria given in \cite{s1}. First, \textit{sparse targeted false data injection attacks} are introduced which provide a strategy for tampering with the measurements from meters in order to build a specific data injection vector. In the second proposed method, called {\it strategic sparse attacks}, the sparse attack vector is constructed by assuming that the attacker has control over only a set of measurements and that the system has secure measurements that cannot be considered in the construction of the attack vector.


Since power grid networks are large scale networks, system monitoring and security control as envisioned for the smart grid are challenging problems. Therefore, decentralized options in which the computational complexity is distributed throughout the network are desirable. For this reason, distributed estimation techniques arise as strong candidates to incorporate adaptability to dynamic network topologies and flexible reconfiguration in case of sub-network faults. Additionally, distributed estimation techniques do not require all network state information to be available to each group, which facilitates operating with limited knowledge about the state of the network.  However, the distributed structure of the networks may lead to critical attacks. For instance, distributed and collective attacks to \textit{active nodes}, which have higher numbers of connections than the rest of the nodes, may cause larger damage to the network (i.e., the group of nodes connected to \textit{active nodes}), because of its scale-free and hierarchical structure \cite{cn,cn1}.

We introduce two distributed attack models that make use of the sparsity of the attack vectors. The first model, \textit{Distributed Sparse Attacks}, assumes that the attacks are directed at clusters of measurements. In this setting, attackers have access to a subset of the measurements observed by the nodes in the cluster. The goal is to achieve a consensus on the design of the attack vectors by iteratively computing them for the measurements observed in each cluster. The second model, \textit{Collective Sparse Attacks}, assumes that the network topology is known by the attackers and may access the measurements observed in the whole network. However, attacks occur in groups, i.e., state variables in the same group are attacked by the same attack vectors.

In addition, we introduce two distributed state vector estimation methods from the perspective of the network operator. The first method, \textit{Distributed State Vector Estimation}, considers the scale-free or hierarchical structure of the network, i.e., the observed measurements are grouped into clusters. Then, local state vector estimates are computed using local measurements by either local network operators or \textit{smart} Phasor Measurement Units (PMUs). Using an iterative message-passing sparse optimization algorithm, each local operator or unit sends the estimated state vectors to centralized network processors, which update the state vector estimates. The second method, called \textit{Collaborative Sparse State Vector Estimation}, assumes that different vector operators estimate a subset of state variables. For instance, different network operators may have expertise or special tools in order to estimate specific state variables and as a result, state vector variables are assumed to be distributed in groups and locally accessed by network operators. In this method, each network operator computes an estimate of the subset of the state variables using local data, and these estimates are then sent to a centralized operator in order to update their values. We analyze the proposed state vector estimation methods for attack detection and identification using a residuals test method in Section VII.

All the optimization problems presented in this paper are solved using the Alternating Direction Method of Multipliers (ADMM) algorithm \cite{admm}. Parameter and stopping criteria selection methods of ADMM are given in \cite{admm} and \cite{sel}.  Moreover, convergence properties of ADMM are analyzed in \cite{admm} and \cite{rate}. 

In the next section, we review the unobservable false data injection and state vector estimation problems. Section III describes centralized sparse attack methods in which the sparse structure of the problem is exploited. In Section IV, we introduce distributed and collaborative state vector estimation methods. We introduce distributed and collective sparse attack models in Section V, and their computational complexity is analyzed in Section VI. We assess the validity of the proposed methods using real-world power systems in Section VII. The paper concludes with Section VIII.

\section{Problem Formulation}

\subsection{System Model}
A review of the problem formulation of false data injection attacks and the state vector estimation problem for attacked systems follows. Consider the DC power flow state acquisition problem \cite{s1} given by
\begin{equation}
\mathbf{z} = \mathbf{H} \mathbf{x} + \mathbf{n},
\label{eq:system_model}
\end{equation}
where $\mathbf{z}\in\mathbb{R} ^ N$ is the vector of measurements, $\mathbf{x} \in \mathbb{R} ^ D$ is the state vector which consists of the voltage phase angles at the buses, $\mathbf{H}\in\mathbb{R}^{N\times D}$ is the measurement Jacobian matrix and $\mathbf{n}\in\mathbb{R} ^ N$ is the measurement noise.

The goal of the network operator is to estimate the state vector and decide whether an attack is present. If the noise is normally distributed with zero mean and independent components, then the following estimator can be employed \cite{s1}:
\begin{equation}
\mathbf{\hat x} = ( \mathbf{H} ^T \mathbf{\Lambda} \mathbf{H}) ^{-1} \mathbf{H} ^T \mathbf{\Lambda} \mathbf{z} \; ,
\end{equation}
where $\mathbf{\Lambda}$ is a diagonal matrix whose diagonal elements are given by $\mathbf{\Lambda} _{ii} = \xi_i ^{-2}$, and $\xi_i ^{2}$ is the variance of the $i$-th measurement for $i=1, ..., N$. The network operator decides that an attack is present if $ \| \mathbf{z} - \mathbf{H} \mathbf{\hat x} \|^2 _2 > \tau$, where $ \| \cdot \| _2$ is the $\ell_2$-norm and $\tau \in \mathbb{R} $ is a given threshold. If $ \| \mathbf{z} - \mathbf{H} \mathbf{\hat x} \| ^2 _2 \leq \tau$ then no attack is declared.

The goal of the attacker is to inject a false data vector $\mathbf{a} \in \mathbb{R} ^ N$ into the measurements without being detected by the operator. Since the attack is performed by changing the values of a subset of all the measurements, the resulting observation model for the operator is
\begin{eqnarray}
\label{eq:attack_obs}
\mathbf{\tilde {z}} = \mathbf{H} \mathbf{x} + \mathbf{a} + \mathbf{n}.
\label{eq:a1}
\end{eqnarray}
Note that for the attack vector $a_i \neq 0$, $\forall i \in \mathcal{A}$, where $\mathcal{A}$ is the set of measurement variable indices with which the attacker tampers. On the other hand, the measurements over which the attacker has no control are the secure variables, $a_i = 0$, $\forall i \in \mathcal{S}$. Note that $\mathcal{S}=\bar{\mathcal{A}}$ where $\bar{\mathcal{(\cdot)}}$ is the set complement operator and $|\mathcal{A} \cup \mathcal{S} |=N$ where $| \cdot |$ denotes set cardinality.

Imposing the constraint $\mathbf{a}= \mathbf{H} \mathbf{c}$, where $\mathbf{c} \in \mathbb{R} ^ D$ is an injected data vector guarantees undetectability via residual tests since it lies in the column space of $\mathbf{H}$ \cite{s1,kosut}. Note that \eqref{eq:a1} can be rewritten as
\begin{equation}
\mathbf{\tilde {z}} = \mathbf{H} { \tilde {\mathbf{x}} } + \mathbf{n},
\label{eq:a2}
\end{equation}
where $\tilde {\mathbf{x}} = \mathbf{x} + \mathbf{c}$ is what an operator unaware of the attack tries to estimate instead of the actual state vector $\mathbf{x}$.

\subsection{Sparsity in the System}

Assuming that an attacker can tamper with a limited number of meters poses the optimization problem in a framework in which the attack vector is sparse. Specifically, if $k$ meters are controlled by the attacker, then $\mathbf{a}$ is at most $k$-sparse, i.e., $ \| \mathbf{a} \|_0 \leq k$, where $ \| \cdot \|_0$ is the $l_0$ norm. In \cite{s1} Liu et al. prove the existence of unobservable attack vectors if $k \geq N-D+1$. Finding the sparsest attack vector that satisfies $\mathbf{a}=\mathbf{H}\mathbf{c}$ is computationally intractable in general. Surprisingly, the solution can be relaxed into a convex optimization problem by using the $\ell_1$-norm as the objective function instead of the $\ell_0$-norm \cite{candes, donoho}. Based on sparse reconstruction techniques, Kim and Poor \cite{kim} provide a greedy approach for the attack vector construction when a subset of the measurements is controlled by the attacker while the remaining measurements are secured.

A second scenario in which the sparsity of the system can be exploited is in the estimation of the state vector. Considering that system states are given by a random process $\lbrace\mathbf{x}_t\rbrace$, the components of the state vector that change \textit{significantly} during an interval $(t,t')$ are defined as
\begin{equation}
\mathcal{X}_{t,t'}=\lbrace i:|\mathbf{x}_t(i)-\mathbf{x}_{t'}(i)|\geq\epsilon\rbrace
\end{equation} 
where $\epsilon$ defines the threshold for change \textit{significance}. That being the case, the operator does not need to estimate all state variables for each time $t$. Assuming that it has a previous state estimate, $\hat{\mathbf{x}}_t$, from time $t$, it can estimate the values that changed above the $\epsilon$ threshold by realizing that
\begin{IEEEeqnarray}{rcl}
\mathbf{y}_{t'}-\mathbf{y}_{t}\; &=& \;\mathbf{H}\mathbf{x}_{t'}-\mathbf{H}\hat{\mathbf{x}}_t+\mathbf{z}_{t'}-\mathbf{z}_{t}\\
\; &=&\; \mathbf{H}(\mathbf{x}_{t'}-\hat{\mathbf{x}}_t)+\mathbf{z}_{t'}-\mathbf{z}_{t}\\
\; &=& \; \mathbf{H}{\boldsymbol \delta}_{t,t'}+\mathbf{z}_{t'}-\mathbf{z}_{t},
\end{IEEEeqnarray}
where ${\boldsymbol \delta}_{t,t'}=\mathbf{x}_{t'}-\hat{\mathbf{x}}_t$ has significantly changed variables given by indices $\mathcal{X}_{t,t'}$. By choosing the significance threshold, $\epsilon$, and the estimation time interval appropriately, $\boldsymbol\delta$ becomes a nearly sparse vector whose $k$ largest components can be recovered by solving a standard compressed sensing problem of the form
\begin{eqnarray}
\begin{matrix} 
& \text{minimize} && \|{\boldsymbol \delta} \| _1 \\
& \text{subject to} && \|\mathbf{y}_{t'}-\mathbf{y}_{t}-\mathbf{H}{\boldsymbol \delta}\|_2^2<\gamma,
\end{matrix}  
\label{eq:l11}
\end{eqnarray}
where $\gamma$ is a regularization parameter.

An additional optimization constraint is imposed by the rank deficiency observed in the measurement Jacobian matrix $\mathbf{H}$ of several IEEE test systems, such as the IEEE 39-Bus \cite{mp}. In Table~\ref{tab:table1}, we show the values of $N$, $D$, rank and R (the ratio of the number of nonzero elements of the entries of $\mathbf{H}$) for test systems. We observe that 9-Bus, 14-Bus, 30-Bus and 39-Bus test systems are rank deficient. Although $\mathbf{H}$ matrices of 57-Bus, 118-Bus, 300-Bus and 3375-Bus test systems are not rank deficient, their R values are greater than those of the rank-deficient matrices. Note that the sparseness increases as the system size increases. Following the sparse nature of the system, \eqref{eq:a1} and \eqref{eq:a2} are formulated as $\ell_1$-norm optimization problems \cite{kim}.
\begin{table}[htbp]
  \centering
  \caption{Rank values of measurement Jacobian matrices of IEEE test systems and the 3375-Bus Polish system plus - winter 2007-08 evening peak system.}
    \begin{tabular}{ccccc}
     \multicolumn{1}{c}{System} & \multicolumn{1}{c}{\textbf{$N$}} & \multicolumn{1}{c}{$D$} & \multicolumn{1}{c}{Rank} & \multicolumn{1}{c}{R} \\
     \hline
     9-Bus & 19 & 9 & 8 & 72.00 \% \\
     14-Bus & 34 & 14 & 3 & 80.25 \% \\
     30-Bus & 71 & 30 & 29 & 90.89 \% \\
     39-Bus & 85 & 39 & 38 & 93.27 \% \\
     57-Bus & 137 & 57 & 57 & 95.22 \% \\
	 118-Bus & 304 & 118 & 118 & 97.64 \% \\	
	 300-Bus & 711 & 300 & 300 & 99.09 \% \\
	 3375-Bus & 7536 & 3375 & 3375 & 99.92 \% \\
    \end{tabular}%
  \label{tab:table1}%
\end{table}%

\section{Centralized Data Sparse Attacks}

\subsection{Sparse Targeted False Data Injection Attacks}

Targeted False Data Injection Attacks consist of attackers constructing false data injection vectors $\mathbf{a}$ corresponding to a given attack vector $\mathbf{c}$. In this section, we introduce two models that employ LASSO and regressor selection algorithms to solve the targeted false data injection problem. 

\subsubsection{Targeted LASSO Attacks}

The sparseness of $\mathbf{c}$ is exploited for targeted false data injection attacks \cite{s1}, where $c_j \in \mathbb{R}$ are fixed and defined by attackers $\forall j \in \mathcal{I}$, for a set $\mathcal{I}$ of indices of the state vector variables that will be attacked. However, $c_j \in \mathbb{R}$ are randomly selected by the attacker according to a probability distribution $\forall j \in \bar{\mathcal{I}}$, where $\bar{\mathcal{I}}$ is the set of off-target variables which are not specifically determined by the attackers. In other words, the attackers do not have control on the variables $c_j \in \bar{\mathcal{I}}$. Note that, $|\mathcal{I} \cup \bar{\mathcal{I}}|=D$.

In order to compute the off-target and targeted attack vectors, we employ the following decomposition  \cite{s1}:
\begin{eqnarray}
\mathbf{a}=\mathbf{H} \mathbf{c} = \sum _{i \in \bar{\mathcal{I}}} c_i \mathbf{h}_{i} + \sum _{j \in \mathcal{I}} c_j \mathbf{h}_{j},
\label{eq:targeted_def}
\end{eqnarray}
where  $\mathbf{h}_{l}$ is the $l$-th column of $\mathbf{H}$. Then, we define a sub-matrix $\mathbf{H}^{\bar{\mathcal{I}}}$ of $\mathbf{H}$ as $\mathbf{H}^{\bar{\mathcal{I}}}= (\mathbf{h}_{j_i}, \cdots , \mathbf{h}_{j_{D-| \bar{\mathcal{I}} |}})$, $\forall j_i \in  \bar{\mathcal{I}}$ and $ 1 \leq i \leq D-| \mathcal{I} |$ \cite{s1} and construct a vector $\mathbf{b}$ in the range space of the attacked measurements, such that $\mathbf{b}= \sum _{j \in \mathcal{I}} \mathbf{h}_{j} \mathbf{c}_{j} $. Using this construction, we relate $\mathbf{b}$ to the measurements $\mathbf{H}^{\bar{\mathcal{I}}}$, such that $\mathbf{P}^{\bar{\mathcal{I}}}=\mathbf{H}^{\bar{\mathcal{I}}} (\mathbf{H^{\bar{\mathcal{I}}}} ^T  \mathbf{H}^{\bar{\mathcal{I}}} ) ^{-1} \mathbf{H^{\bar{\mathcal{I}}}} ^T  $, $\mathbf{B}^{\bar{\mathcal{I}}} = \mathbf{P}^{\bar{\mathcal{I}}} - \mathbf{I}$ and $\mathbf{y}=\mathbf{B}^{\bar{\mathcal{I}}} \mathbf{b}$ \cite{s1}. Therefore, we can compute $\mathbf{a}$ by solving $\mathbf{y}=\mathbf{B}^{\bar{\mathcal{I}}} \mathbf{a}$ \cite{s1}. 

We assume that given an attack vector $\mathbf{a}$, the attack strategy of an attacker is to find a sparse $\mathbf{a}$, such that $\mathbf{y}=\mathbf{B}^{\bar{\mathcal{I}}} \mathbf{a}$. Then, using $l_1$ relaxations for sparse vector estimation \cite{candes,donoho,lasso}, we introduce the following optimization problem to model the sparse false data injection attack:
\begin{eqnarray}
\begin{matrix} 
& \text{minimize} && \| \mathbf{a} \| _1 \\
& \text{subject to} && \mathbf{y}=\mathbf{B}^{\bar{\mathcal{I}}} \mathbf{a} .
\end{matrix}  
\label{eq:l1}
\end{eqnarray}
\eqref{eq:l1} is called basis pursuit and can be employed to find a sparse solution vector $\mathbf{c}$ \cite{candes,donoho,lasso}. In order to solve the optimization problem above using ADMM \cite{admm}, \eqref{eq:l1} is formulated as 

\begin{eqnarray}
\begin{matrix} 
& \text{minimize} && I(\mathbf{a}) + \| \bm{\beta} \| _1 \\
& \text{subject to} && \mathbf{a} - \bm{\beta}=\mathbf{0},
\end{matrix}  
\label{eq:l1_admm}
\end{eqnarray}
where $I(\mathbf{a}) $ is the indicator function for $\{ \mathbf{a} \in \mathbb{R} ^ N : \mathbf{y}= \mathbf{B}^{\bar{\mathcal{I}}} \mathbf{a} \}$ and $\bm{\beta} \in \mathbb{R} ^ N$ is the optimization variable. The sparsity of $\mathbf{a}$ is governed by $\| \bm{\beta} \| _1$ using a scalar real number $\lambda > 0$, which is a regularization parameter. Moreover, in order to reduce the probability of the attack being detected,  $\| \mathbf{y} - \mathbf{B}^{\bar{\mathcal{I}}} \mathbf{{a}} \| ^2 _2$ can be used as a cost function which results in the optimization problem
\begin{eqnarray}
\begin{matrix} 
& \text{minimize} && \| \mathbf{y} - \mathbf{B}^{\bar{\mathcal{I}}} \mathbf{{a}} \| ^2 _2 + \lambda \| \bm{\beta} \| _1 \\
& \text{subject to} && \mathbf{a} - \bm{\beta}=0.
\end{matrix}  
\label{eq:l12_admm}
\end{eqnarray}
Problem \eqref{eq:l12_admm} is a \textit{LASSO} optimization \cite{lasso} and can be solved via ADMM \cite{admm} as follows:

\begin{Alg}[LASSO via ADMM]
\label{alg:admm1}
\end{Alg}
\begin{itemize}
\item INPUT:
\begin{itemize}
\item Projection matrix defined by secure set  $\mathbf{B}^{{\bar{\mathcal{I}}}}$
\item Projected vector containing injected data $\mathbf{y}$ 
\item Penalty parameter $\rho$
\item Maximum number of iterations $t'$
\end{itemize}
\item OUTPUT:
\begin{itemize}
\item Attack vector candidate $\mathbf{a}\stackrel{\sf def}{=}\mathbf{a}^t$
\end{itemize}
\item PROCEDURE:
\begin{enumerate}
\item Initialize $t=0$, $\bm{\beta}=\mathbf{0}$ and $\mathbf{u}=\mathbf{0}$
\item Compute \textit{ridge regression}  with penalty parameter $\rho$:
 \begin{equation}
 \mathbf{a} ^{t+1}= \Big ( (\mathbf{B}^{\bar{\mathcal{I}}}) ^T  \mathbf{B}^{\bar{\mathcal{I}}} + \rho I \Big ) ^{-1} \Big ( (\mathbf{B}^{\bar{\mathcal{I}}}) ^T  \mathbf{y} + \rho ( \bm{\beta} ^t - \mathbf{u} ^t) \Big )
 \label{eq:a_step}
 \end{equation}
 \item Perform \textit{soft thresholding} defined by proximity operator $\Pi _{ \kappa} (\phi)=(\phi -  \kappa)_+ - (-\phi -  \kappa)_+$, where $(\phi)_+ ={\sf max}(\phi,0)$:
 \begin{equation}
 \bm{\beta} ^{t+1} = \Pi _{ \lambda / \rho } (\mathbf{u} ^t+ \mathbf{a} ^{t+1} ) 
 \end{equation}
 \item Update: 
 \begin{equation}
 \mathbf{u} ^{t+1} = \mathbf{u} ^{t} + \mathbf{a} ^{t+1} - \bm{\beta} ^{t+1}
 \end{equation}
 \item Return to step 2 if a stopping criterion is not satisfied and $t<t'$
\end{enumerate}

\end{itemize}

\subsubsection{Selective Targeted Attacks}
The previous approach provides an implicit control of the sparsity of $\mathbf{a}$ using parameter $\lambda$. In the following, the sparsity is explicitly controlled by introducing the constraint $ \| \mathbf{a} \|_0 \leq k$ in \eqref{eq:l1} in the optimization problem as
\begin{eqnarray}
\begin{matrix} 
& \text{minimize} && \| \mathbf{y} - \mathbf{B}^{\bar{\mathcal{I}}} \mathbf{{a}} \| ^2 _2  \\
& \text{subject to} && \| \mathbf{a} \| _0 \leq k \; .
\end{matrix}  
\label{eq:s1}
\end{eqnarray}
This optimization problem can also be solved via ADMM with a minor modification of Algorithm \ref{alg:admm1}. Specifically, hard-thresholding $\Pi ^* _{ \lambda / \rho } ( \mathbf{a} ^{t+1} + \mathbf{u} ^t)$ is employed in the update of $\bm{\beta}$, such that $k$ largest magnitude elements of $\mathbf{u} ^t+\mathbf{a} ^{t+1}$ are kept and zeros are assigned to the remaining elements \cite{admm}.

\subsection{Strategic Sparse Attacks}

In this section, we propose two algorithms to compute the attack vector $\mathbf{c}$ based on the formulations of \textit{LASSO Attacks} and \textit{Selective Sparse Attacks} for the strategic sparse attack model case introduced by Kim and Poor \cite{kim}. To this end, we first redefine the sparse data injection attack problem for ADMM. Then, we solve the optimization problems using LASSO and Regressor Selection algorithms.  

\subsubsection{Strategic Sparse Attacks with LASSO}

In Strategic Sparse Attacks, a row-wise decomposition of the Jacobian measurement matrix is employed based on the set $\mathcal{A}$ of attacked measurement indices denoting the meters to which an attacker has access, and the set $\mathcal{S}$ of secure measurement indices, i.e., the indices of meters which cannot be tampered by an attacker. Specifically, a sub-matrix $\mathbb{H}^{\mathcal{S}}= (\mathbf{H}_{j_i,:}, \cdots \mathbf{H}_{j_{N-|\mathcal{S}|,:}} )$, $\forall j_i \in  \mathcal{S}$, of $\mathbf{H}$ is constructed in order to represent the secure measurements, where $\mathbf{H}_{j_i,:}$ is the $j_i$-th row of $\mathbf{H}$, such that $\mathbb{H}^{\mathcal{S}} \mathbf{c}=\mathbf{0}$. Similarly, sub-matrix $\mathbb{H}^{\mathcal{A}}$ is defined for attacked measurements. As a result, the attacker's strategy is defined to find a solution $\mathbf{c}$ to the following optimization problem:
\begin{eqnarray}
\begin{matrix} 
& \text{minimize}  & \| \mathbb{H}^\mathcal{A} \mathbf{c} \| _0 \\
& \text{subject to} & \mathbb{H}^\mathcal{S} \mathbf{c} = \mathbf{0} \; , \\
					&& \| \mathbf{c} \| _{ \infty}  \geq \psi \; ,
\end{matrix}  
\label{eq:sl00}
\end{eqnarray}
where $\psi \geq 0$ is a given constant \cite{kim}.

Define $\mathbf{h}_i$ as the $i$-th column vector of $\mathbf{H}$, the sub-matrix $\mathbb{H}_i\in\mathbb{R}^{N\times (D-1)}$ formed by removing the $i$-th column of $\mathbf{H}$, and $\boldsymbol{\sigma}_i \in \mathbb{R} ^{D-1}$ formed by removing the $i$-th variable of $\mathbf{c}$. Following these definitions, the strategic sparse attack is defined as
\begin{eqnarray}
\begin{matrix} 
& \text{minimize}  && \| \mathbb{H} ^\mathcal{A}_i \boldsymbol{\sigma}_i + \mathbf{h}^\mathcal{A}_{i} \| _1 \\
& \text{subject to} && \mathbb{H}^\mathcal{S} _i \boldsymbol{\sigma}_i + \mathbf{h} ^\mathcal{S}_{i} = \mathbf{0} \; . 
\end{matrix}  
\label{eq:sl01}
\end{eqnarray}

Since $\mathbf{H}$ and $\mathbf{c}$ are sparse, it follows that the problem can be reformulated as
\begin{eqnarray}
\begin{matrix} 
& \text{minimize}  && \| \boldsymbol{\sigma}_i \| _1 \\
& \text{subject to} && \mathbb{H} ^\mathcal{S} _i \boldsymbol{\sigma}_i + \mathbf{h} ^\mathcal{S} _{i} = \mathbf{0} \; . 
\end{matrix}  
\label{eq:sl1}
\end{eqnarray}

Since \eqref{eq:sl1} is a LASSO problem, we reformulate \eqref{eq:sl1} as an ADMM optimization problem as follows:
\begin{eqnarray}
\begin{matrix} 
& \text{minimize} && \| \mathbb{H} ^\mathcal{S} _i \boldsymbol{\sigma}_i + \mathbf{h} ^\mathcal{S} _{i} \| ^2 _2 + \lambda \| \boldsymbol{\theta}_i \| _1 \\
& \text{subject to} && \boldsymbol{\sigma}_i - \boldsymbol{\theta}_i=\mathbf{0}\; ,
\end{matrix}  
\label{eq:sl1_admm}
\end{eqnarray}
where $\boldsymbol{\theta}_i \in \mathbb{R}^{D-1}$ is the optimization variable. In order to solve \eqref{eq:sl1_admm}, Algorithm \ref{alg:admm1} can be employed with inputs $\mathbb{H} ^\mathcal{S} _i$, $\mathbf{y}=1$ and $\theta_i$. This procedure is repeated for $i=1,...,D$ in order to compute $\mathbf{c}=(\mathbf{c}_1,...,\mathbf{c}_D)$.

\subsubsection{Selective Strategic Sparse Attacks}
As discussed in the previous section, the sparsity of $\mathbf{c}_i$ can be bounded explicitly by converting \eqref{eq:sl1} to the equivalent regressor selection problem given by
\begin{eqnarray}
\begin{matrix} 
& \text{minimize} && \| \mathbb{H} ^\mathcal{S}_i \boldsymbol{\sigma}_i + \mathbf{h} ^\mathcal{S}_{i} \| ^2 _2 \\
& \text{subject to} && \| \boldsymbol{\sigma}_i \| _0 \leq k \; .
\end{matrix}  
\label{eq:ss1}
\end{eqnarray}
In this formalism, we relax the constraint in $\mathbb{H}^\mathcal{S} \mathbf{c} = \mathbf{0}$ and introduce a sparsity constraint in the construction of attack vectors $\mathbf{c}$, such that we compute an attack vector with at most $k$ non-zero elements. The solution to \eqref{eq:ss1} is the same as the one proposed for \eqref{eq:sl1_admm} except for substituting the soft thresholding operator in step 3 by a hard thresholding.

\subsection{Computational Complexity of Centralized Sparse Attacks}

The optimization problems of the centralized sparse attacks are solved using Algorithm 1. The computational complexity of the algorithm is dominated by the attack vector update step in \eqref{eq:a_step} which solves a ridge regression problem \cite{ridge}. Therefore, the computational complexity of the algorithm is $\Upsilon_1 \in O(t' \alpha^3)$, where 
\begin{enumerate}
\item $\alpha=\min(N, |\bar{\mathcal{I}}|) $ for targeted attacks given in Section III.A, and
\item $\alpha=\min(N, D-1) $, for strategic attacks given in Section III.B.
\end{enumerate}
Note that, the computational complexity of the algorithm is increased by an additional term $D$ (the dimension of the attack vector) to $O(t'D\alpha^3)$ for strategic attacks, since the algorithm is implemented $D$ times. 

In the implementation, the running or iteration time $t'$ can be relaxed by using performance based early stopping criteria as suggested in \cite{admm}.

\section{Distributed and Collaborative Sparse State Vector Estimation}

A sparse state vector estimation model, called  \textit{Distributed Sparse State Vector Estimation}, is first introduced in order to estimate the state vectors under attack on the network measurements using an instance distributed \textit{LASSO} algorithm. In the second model, called \textit{Collaborative Sparse State Vector Estimation}, we assume that the topological information of the network and the measurements are available to the network operator. However, the network operator can choose to process different groups of state vectors using the \textit{Group LASSO} algorithm. We solve the optimization problems using the ADMM algorithm.

\subsection{ Distributed Sparse State Vector Estimation }

Measurements are distributed in the network and usually form clusters following the topological properties of the network. Additionally, observation vectors and measurement matrices are partitioned into $G$ blocks denoted by $\mathcal{G}_i$ with $|\mathcal{G}_i|=N_i$ for $i=1, \ldots, G$. As a result, the attacks can also be partitioned. Taking this into account, \eqref{eq:attack_obs} can be rewritten as

\begin{eqnarray}
\begin{bmatrix} 
\mathbf{\tilde{z}}_{1} \\ 
\vdots \\
\mathbf{\tilde{z}}_{G} \\

\end{bmatrix}  
= 
\begin{bmatrix} 
\mathbf{H}_{1} \\ 
\vdots \\
\mathbf{H}_{G} \\

\end{bmatrix}
\mathbf{x}
+
\begin{bmatrix} 
\mathbf{a}_{1} \\ 
\vdots \\
\mathbf{a}_{G} \\
\end{bmatrix}
+
\begin{bmatrix} 
\mathbf{n}_{1} \\ 
\vdots \\
\mathbf{n}_{G} \\
\end{bmatrix} \;,
\label{eq:dve1}
\end{eqnarray}  
where $\tilde{\mathbf{z}}_{i} \in \mathbb{R} ^ {N_i}$ is the measurement observed in the $i$-th cluster of nodes through measurement matrix $\mathbf{H_{i}}\in\mathbb{R}^{N_i \times D}$ and noise $\mathbf{n}_{i} \in \mathbb{R} ^ {N_i}$, and which is under attack $\mathbf{a}_i \in \mathbb{R} ^ {N_i}$ with $i=1,...,G$.  For each cluster, we consider the penalty function 
\begin{equation}
f_i= \| \tilde{\mathbf{z}}_{i} - \mathbf{H}_i \tilde{\mathbf{x}}_{i} \| ^2 _2 \;,
\end{equation}
where $\tilde{\mathbf{x}}_i$ is the state vector estimated at cluster $i$. Note that 
\[
f\equiv\| \tilde{\mathbf{z}} - \mathbf{H} \tilde{\mathbf{x}} \| ^2 _2=\sum ^G _{i=1} f_i.
\]
Therefore, we can write the distributed optimization problem in the following way:
\begin{eqnarray}
& \text{minimize} & \sum ^G _{i=1} f_i + g(\bm{\beta}) \\
& \text{subject to} & \mathbf{ \tilde{\mathbf{x}}}_i - \bm{\beta} =\mathbf{0}, \quad i=1,...,G \;,
\label{eq:da2}
\end{eqnarray}
where $\bm{\beta} \in \mathbb{R} ^ {D}$ is the optimization variable, $g(\bm{\beta})=\lambda \| \bm{\beta} \| _1$ is the regularization function and $\lambda \in \mathbb{R}$ is the regularization parameter, which controls the sparsity of the solution vector. Since network operators accessing the local data should agree on the estimated state vector, we introduce a \textit{consensus} constraint in \eqref{eq:da2}. In other words, \eqref{eq:da2} is considered as a global consensus problem in which $\bm{\beta}$ is used as the \textit{global} optimization variable.

We solve \eqref{eq:da2} using an ADMM implementation as described in the following algorithm.

\begin{Alg}[Distributed Estimation via ADMM]
\label{alg:dist_est}
\end{Alg}
\begin{itemize}
\item INPUT:
\begin{itemize}
\item Projection matrix $\mathbf{H}$
\item State measurements $\mathbf{\tilde{z}}$ 
\item Set of clusters $\lbrace \mathcal{G}_i\rbrace_{i=1}^G$
\item Penalty parameter $\rho$
\item Maximum number of iterations $t'$
\end{itemize}
\item OUTPUT:
\begin{itemize}
\item Estimated state vector $\tilde{\mathbf{x}} \equiv \tilde{\mathbf{x}}^t$
\end{itemize}
\item PROCEDURE:
\begin{enumerate}
\item Initialize $t=0$, $\bm{\beta}^0=\mathbf{0}$, $\mathbf{u}^0=\mathbf{0}$.
\item For $i=1,\ldots, G$ compute the Tikhonov-regularized least squares estimate with penalty parameter $\rho$ given by
 \begin{equation}
\tilde{\mathbf{x}} ^{t+1} _i = \left(\mathbf{H} ^T _i \mathbf{H}_i + \rho\mathbf{I} \right) ^{-1}\left(\mathbf{H} ^T _i \tilde{\mathbf{z}}_{i} + \rho ( \bm{\beta} ^t - \mathbf{u} ^t _i )\right).
\label{eq:x_update_est}
 \end{equation}
 \item Perform a \textit{soft thresholding} given by
  \begin{equation}
\bm{ \beta} ^{t+1} = \Pi _{ \frac{\lambda}{\rho G}��}\left( \frac{1}{G} \sum ^G _{i=1} \left (\tilde{\mathbf{x}} ^{t+1} _i+ \mathbf{ u} ^{t} _i\right )\right),
 \end{equation}
where the $\ell_1$ proximity operator is defined as
  \begin{equation}
 \Pi _{ \kappa} (\phi)=(\phi -  \kappa)_+ - (-\phi -  \kappa)_+
 \end{equation}
and $(\phi)_+ ={\sf max}(\phi,0)$.

 \item For $i=1,\ldots, G$ update
 \begin{equation}
\mathbf{ u} ^{t+1} _i = \mathbf{ u} ^{t} _i + \tilde{\mathbf{x}} ^{t+1} _i - \bm{\beta} ^{t+1}.
\end{equation}
\item Return to step 2 if the halting criterion is not satisfied and $t<t'$.
\end{enumerate}
\end{itemize}
Note that $\mathbf{H} _i$ is a sparse matrix or vector (depending on $\mathcal{G}_i$). Still, $(\mathbf{H} ^T _i \mathbf{H}_i + \rho \mathbf{I} )$ is invertible since $\rho > 0$. 

Algorithm and optimization variables are initialized in the first step of the algorithm. In the second step, each network operator computes a local estimate using Tikhonov-regularized least squares \cite{tikhonov,tik_power}. Then the local estimates are \textit{gathered} to update the global variable $\bm{\beta}$ in the third step. Finally, the updated $\bm{\beta}$ is distributed or \textit{broadcast} to the clusters to update the dual variables  $\mathbf{u}_i$, $\forall i=1, \ldots, G$, in the fourth step, and the halting criterion is checked in the last step.

\subsection{ Collaborative Sparse State Vector Estimation }
In the distributed sparse attacks scenario, measurements are assumed to be distributed across clusters and operators have access only to local measurements. Alternatively, when collective sparse attacks are considered, operators know the whole topology of the network and the Jacobian measurement matrix $\mathbf{H}$. However, in a distributed framework operators observe a subset of state vector variables, i.e., each operator may observe different groups of buses.

In this setting, the observation model \eqref{eq:a1} can be rewritten as
\begin{eqnarray}
\mathbf{\tilde{z}}
= 
\begin{bmatrix} 
\mathbf{\hat{H}}_{1} 
\cdots 
\mathbf{\hat{H}}_{G} 
\end{bmatrix}
\begin{bmatrix} 
\mathbf{x}_{1} \\
\vdots \\
\mathbf{x}_{G} 
\end{bmatrix}
+
\mathbf{a} 
+
\mathbf{n} \;,
\label{eq:cve1}
\end{eqnarray}  
where $\tilde{\mathbf{z}} \in \mathbb{R} ^ {N}$ is the measurement vector, $\mathbf{x}_{i} \in \mathbb{R} ^ {D_i}$ is the state vector, $\mathbf{n} \in \mathbb{R} ^ {N}$ is a noise vector and $\mathbf{\hat{H}_{i}}\in\mathbb{R}^{N\times D_i}$ is the Jacobian measurement submatrix formed by selecting the columns given by the indices of the subset of state variables assigned to cluster $i$. Given this structure, the optimization problem can be stated as
\begin{eqnarray}
\text{minimize} & \| \mathbf{H} \mathbf{\tilde{\mathbf{x}}} - \mathbf{\tilde{\mathbf{z}}} \| ^2 _2 + \lambda\displaystyle \sum ^G _{i=1} \| \tilde{\mathbf{x}} _i \| _2.
\label{eq:cd1}
\end{eqnarray}
By introducing an optimization variable $ \mathbf{v} \in \mathbb{R} ^ {D}$, it follows that
\begin{eqnarray}
\text{minimize} & \| \mathbf{H} \mathbf{v} - \tilde{\mathbf{z}} \| ^2 _2 + \lambda \displaystyle\sum ^G _{i=1} \| \tilde{\mathbf{x}}_i \| _2 \\
\text{subject to} & \quad\tilde{\mathbf{x}}_i - \mathbf{\hat {v}}_i =\mathbf{0}, \quad i=1,...,G \;,
\label{eq:cd2}
\end{eqnarray}
is equivalent to \eqref{eq:cd1}, where $ \mathbf{\hat {v}}_i$ is the estimate of $ \mathbf{v}$ for $\tilde{\mathbf{x}}_i$ \cite{admm}. In order to solve \eqref{eq:cd2}, the proposed ADMM implementation is described below.

\begin{Alg}[Collaborative Estimation via ADMM]
\label{alg:admm}
\end{Alg}
\begin{itemize}
\item INPUT:
\begin{itemize}
\item Projection matrix $\mathbf{H}$
\item State measurements $\mathbf{\tilde{z}}$ 
\item Set of clusters $\lbrace \mathcal{G}_i\rbrace_{i=1}^G$
\item Penalty parameter $\rho$
\item Maximum number of iterations $t'$
\end{itemize}
\item OUTPUT:
\begin{itemize}
\item Estimated state vector $\tilde{\mathbf{x}} \equiv \tilde{\mathbf{x}}^t$
\end{itemize}
\item PROCEDURE:
\begin{enumerate}
\item Initialize $t=0$, $\bm{\beta}^0=\mathbf{0}$, $\mathbf{v}^0=\mathbf{0}$, $\bm{\theta}^0=\bm{0}$ and $\tilde{\mathbf{x}}^0=\mathbf{0}$.
\item For $i=1,\ldots, G$ compute

 \begin{equation}
\tilde{\mathbf{x}}^{t+1} _i = \underset{ \tilde{\mathbf{x}} _i }{\mathrm{argmin}} \left( \rho \| \bm{\theta}^t_i \| ^2 _2 + 
 \lambda \| \tilde{\mathbf{x}} _i \| _2  \right ),
 \label{eq:x_step_algo3}
 \end{equation}
where $\bm{\theta}^t_i=\mathbf{\hat{H}}_i\left (\tilde{\mathbf{x}}_i - \tilde{\mathbf{x}} ^t _i \right )- \bar{\mathbf{v}} ^t +  \overline{\mathbf{H} \tilde{\mathbf{x}} } ^{t}+ \mathbf{u}^t$ and \\ $ \overline{\mathbf{H} \tilde{\mathbf{x}} } ^{t}=\frac{1}{G} \sum ^G _{i=1} \mathbf{\hat{H}} _i \tilde{\mathbf{x}} _i ^{t}.$
 
 \item Update
 \begin{IEEEeqnarray}{rCl}  
\bar{\mathbf{ v}} ^{t+1}& = &\frac{1}{G + \rho } ( \tilde{\mathbf{z}} + \rho \overline{\mathbf{H} \tilde{\mathbf{x}} } ^{t+1} + \rho \mathbf{u} ^t),\\
\mathbf{ u} ^{t+1} &=& \mathbf{ u} ^{t} + \overline{\mathbf{H} \tilde{\mathbf{x}} } ^{t+1} - \bar{\mathbf{ v}} ^{t+1}.
\end{IEEEeqnarray}
\item Return to step 2 if the halting criterion is not satisfied and $t<t'$.
\end{enumerate}
\end{itemize}

\section{Distributed and Collective Sparse Attacks}

In Section III, we introduced centralized sparse attack methods. In this section, two distributed attack models are proposed in order to employ sparse attacks in a distributed framework. For this purpose, the structure of the measurements and the attack vectors is redefined, followed by a formulation of the false data injection problem as a distributed sparse optimization problem.

The proposed distributed attack models are motivated by two distributed attack scenarios.
\begin{enumerate}

\item In \textit{Distributed Sparse Attacks}, \textit{measurements} are assumed to be distributed in the network and may form clusters following the topological properties of the network. Therefore, different attackers located in different clusters can construct attack vectors by just analyzing the local measurements observed in the clusters. 

\item \textit{Collective Sparse Attacks} model assumes that attackers know the whole topology of the network and the Jacobian measurement matrix $\mathbf{H}$. However, in this case the attacks are directed at a group of \textit{state vector variables} distributed in the network, \textit{i.e.}, each attack injects false data into the state vector variables of the corresponding cluster.
\end{enumerate}

Although linear sparse attacks are considered for the implementation of distributed attacks in this work, the proposed parallelization and distributed processing strategies can be used as design patterns for developing distributed sparse targeted false data injection attacks and strategic sparse attacks. 

\subsection{Distributed Sparse Attacks}

A linear sparse attack model is considered, in which given an attack vector, $\mathbf{a}$, the attack strategy is to find a sparse injection vector, $\mathbf{c}$, such that $\mathbf{a}=\mathbf{H} \mathbf{c}$ \cite{s1}. Using an $\ell_1$ relaxation for sparse vector estimation \cite{candes,donoho,lasso}, the following optimization problem is considered:
\begin{eqnarray}
\begin{matrix} 
& \text{minimize} && \| \mathbf{c} \| _1 \\
& \text{subject to} && \mathbf{a}=\mathbf{H} \mathbf{c}. 
\end{matrix}  
\label{eq:dl1}
\end{eqnarray}

As noted before, measurements are assumed to be distributed in the network and may form clusters following the topological properties of the network. Similar to the partitioning presented in the previous section, the Jacobian measurement matrix is partitioned into $G$ number of submatrices, which results in a partitioning of the attack vector given by
\begin{eqnarray}
\begin{bmatrix} 
\mathbf{a}_{1} \\ 
\vdots \\
\mathbf{a}_{G} \\
\end{bmatrix}  
= 
\begin{bmatrix} 
\mathbf{H}_{1} \\ 
\vdots \\
\mathbf{H}_{G} \\
\end{bmatrix}
\mathbf{c}.
\label{eq:ds1}
\end{eqnarray} 
Note that, \eqref{eq:ds1} can also be employed in the distributed false data vector construction described in \eqref{eq:dve1}. 

In order to solve \eqref{eq:dl1} in the distributed form set by \eqref{eq:ds1} using a distributed optimization algorithm, the loss function is assumed to be separable, such that 
\begin{equation}
 f_i= \| \mathbf{a}_i - \mathbf{H}_i \mathbf{{c}}_i \| ^2 _2.
 \end{equation} 
Note that, $ \sum ^G _{i=1} f_i =f$. Moreover, the optimization problem \eqref{eq:dl1} is assumed to be feasible \cite{admm}. Therefore, the distributed optimization problem for \eqref{eq:dl1} can be reformulated as
\begin{eqnarray}
\text{minimize}& \sum ^G _{i=1} f_i + g(\boldsymbol{\phi}) \\
\text{subject to}&\quad\quad\quad\quad\quad\mathbf{ {c}}_i - \boldsymbol{\phi} =\mathbf{0},\quad i=1,...,G \; ,
\label{eq:dl2}
\end{eqnarray}
where $\boldsymbol{\phi} \in \mathbb{R} ^ {N_i}$ is the optimization variable, $g(\boldsymbol{\phi})=\lambda \| \boldsymbol{\phi} \| _1$ is the regularization function and $\lambda \in \mathbb{R}$ is the regularization parameter which controls the sparsity of the solution vector. Interestingly, the optimization problem \eqref{eq:dl2} is the same as the one posed in \eqref{eq:da2} and therefore, Algorithm \ref{alg:dist_est} can be used to solve it.

\subsection{Collective Sparse Attacks}

The collective sparse attacks model assumes that attackers know the whole topology of the network and the Jacobian measurement matrix $\mathbf{H}$. However, in this case the attacks are directed at a group of state vector variables, \textit{i.e.}, each attack injects false data into the state vector variables of the corresponding cluster. Within this setting, \eqref{eq:dl1} can be rearranged as
\begin{eqnarray}
\mathbf{a}
= 
\begin{bmatrix} 
\mathbf{\hat{H}}_{1} 
\cdots 
\mathbf{\hat{H}}_{G} 
\end{bmatrix}
\begin{bmatrix} 
\mathbf{c}_{1} \\
\vdots \\
\mathbf{c}_{G} 
\end{bmatrix} \; ,
\label{eq:cs1}
\end{eqnarray}
where the injection vector $\mathbf{c}_i \in \mathbb{R} ^ {D_i}$ is computed by the $i$-th attacker using Jacobian measurement submatrix $\mathbf{\hat{H}}_i$ for $i=1,...,G$. Following this decomposition, the optimization can be posed as
\begin{eqnarray}
\text{minimize} &\| \mathbf{H} \mathbf{c} - \mathbf{a} \| ^2 _2 + \lambda\displaystyle \sum ^G _{i=1} \| \mathbf{c}_i \| _2.
\label{eq:g1}
\end{eqnarray}
Introducing the optimization variables, $ \boldsymbol{\psi} \in \mathbb{R} ^ {D}$, yields a new formulation 
\begin{eqnarray}
\text{minimize} & \| \mathbf{H} \boldsymbol{\psi} - \mathbf{a} \| ^2 _2 + \lambda \displaystyle\sum ^G _{i=1} \| \mathbf{c}_i \| _2 \\
\text{subject to} & \mathbf{ {c}}_i - \boldsymbol{\hat {\psi}}_i =\mathbf{0} \;, i=1,...,G \; ,
\label{eq:g2}
\end{eqnarray}
where $ \boldsymbol{\hat {\psi}}_i$ is the estimate of $ \boldsymbol{\psi}$ for $\mathbf{ {c}}_i$ \cite{admm}. 
In the same fashion as with the previous problem, the optimization problem \eqref{eq:g2} is the same as \eqref{eq:cd1} and therefore, Algorithm \ref{alg:admm} can be used to solve it.

\section{Computational Complexity of Distributed Algorithms}

We solve the distributed optimization problems using two main approaches, namely \textit{measurement distributed} and \textit{attribute distributed} optimization as given in Algorithm 2 and Algorithm 3 respectively. In the measurement distributed approach, we assume that the measurements are distributed and the local solutions of the optimization algorithms are computed in the clusters. In the attribute distributed approach, we assume that the state or attack vector variables are distributed and local estimates are computed in the clusters.

If we ignore communication times required to \textit{gather} and \textit{broadcast} the locally estimated vectors $\tilde{\mathbf{x}}_i$ and local variables $\mathbf{u}_i$, then the computational complexity of Algorithm 2 is dominated by the $\tilde{\mathbf{x}}_i$-update step in \eqref{eq:x_update_est}, $\forall i=1, \ldots, G$. Since the partitioned Jacobian matrix $\mathbf{H}_i$ is used in \eqref{eq:x_update_est}, the computational complexity of \eqref{eq:x_update_est} is $O(\alpha_i^3)$, where $\alpha_i=\min(N_i,D)$ in each cluster $\mathcal{G}_i$. Then, the complexity of Algorithm 2 is $ \Upsilon_2 \in \max \Big( t'O(\alpha_1^3), \ldots, t'O(\alpha_G^3) \Big)$, since a central processor which employs the third step of the algorithm should wait to gather all the local estimates from the processors in the clusters. If we define the maximum communication complexity of gathering the local data as $\Upsilon_g$ and that of broadcasting as $\Upsilon_b$, then the complexity of Algorithm 2 is increased to $ \Upsilon_2 +\Upsilon_g+\Upsilon_b$.

Similarly, $G$ parallel regularized least squares problems are solved in $G_i$ variables in the $\tilde{\mathbf{x}}_i$-update step \eqref{eq:x_step_algo3} of Algorithm 3.  Since data partitioning by attribute is employed, the computational complexity of \eqref{eq:x_step_algo3} is $O(\alpha_i^3)$, where $\alpha_i= \min(N,D_i)$. Similarly, the complexity of Algorithm 3 is $ \Upsilon_3 \in \max \Big( t'O(\alpha_1^3), \ldots, t'O(\alpha_G^3) \Big)$, and the communication cost increases the complexity to $ \Upsilon_3 +\Upsilon_g+\Upsilon_b$.

In the implementations, several practical tricks such as caching can be used to decrease the computational complexity of the local optimization algorithms \eqref{eq:x_update_est} and \eqref{eq:x_step_algo3}. For further details, please refer to \cite{admm}.

\section{Numerical Results}

In this section, the validity of the proposed algorithms is assessed by numerically evaluating the performance of the algorithms for IEEE 9-Bus, IEEE 30-Bus, IEEE 57-Bus and IEEE 118-Bus test systems \cite{mp}. For each data point 100 realizations are simulated. For all simulation results, $\lambda$ is fixed as \cite{admm}
\begin{equation}
\lambda = C \lambda_{max},
\label{eq:lambda}
\end{equation} 
where $C$ is a constant, $\lambda_{max}= \| \mathbf{H} \tilde{\mathbf{z}} \| _{\infty} $ for distributed sparse state vector estimation methods, and $\lambda_{max}= \| \mathbf{H} \mathbf{a} \| _{\infty} $ for distributed attack models. In addition, $\lambda_{max}$ can be considered as a critical value of the regularization parameter $\lambda$ above which the estimated state and attack vectors take zero values. Consequently, $C$ determines the \textit{sparsity} of the solutions of the optimization problems and the number of iterations required to obtain the solutions, \textit{i.e.} the estimated state and attack vectors. For that reason, an \textit{optimal} $\hat{\lambda}$ or $\hat{C}$ is computed by analyzing the solution (or regularization) path of the optimization algorithms using a given training dataset. A detailed analysis of the impact of $C$ on the number of algorithm iterations required to obtain an optimal solution is given in \cite{admm} for ADMM implementations of LASSO type algorithms. We choose the penalty parameter as $\rho=1$, the absolute tolerance as $10^{-4}$, the relative tolerance as $10^{-2}$ and set the maximum number of iterations $t'=10000$.

In the experiments, it is assumed that the attacker has access to $k$ measurements. For each realization, a $k$-sparse attack vector, $\mathbf{a}$, is randomly generated by selecting the non-zero indices following a uniform distribution and Gaussian distributed amplitudes with the same mean and variance values as $\mathbf{z}$. For distributed instances of the problem, the number of clusters, $G$, is uniformly distributed from the set of all prime divisors of $N$. On the other hand, for the collaborative instances, $G$ is chosen from the set of all prime divisors of $D$.

\subsection{Results for Centralized Data Injection Attacks}

In order to assess the performance, the following parameters are computed in the simulations:
\begin{enumerate}
\item $Pr(\mathbf{\hat{a}}_i \neq 0 , \mathbf{a}_i \neq 0 ) $ or simply $Pr(\mathbf{\hat{a}'}_i, \mathbf{a'}_i)$, which is the probability of correctly constructing an attack variable $\mathbf{\hat{a}}_i \neq 0$ of a false data injection vector $\mathbf{a}$.
\item $Pr(\mathbf{\hat{a}}_i = 0 , \mathbf{a}_i = 0 ) $ or simply $Pr(\mathbf{\hat{a}}_i, \mathbf{a}_i)$, which is the probability of correctly constructing a secure variable $\mathbf{\hat{a}}_i = 0$ of a false data injection vector $\mathbf{a}$.
\end{enumerate}
Since $Pr(\mathbf{\hat{a}'}_i, \mathbf{a'}_i) + Pr(\mathbf{\hat{a}}_i, \mathbf{a'}_i)=1$ and $Pr(\mathbf{\hat{a}}_i, \mathbf{a}_i) + Pr(\mathbf{\hat{a}'}_i, \mathbf{a}_i)=1$, probabilities of incorrect constructions can be computed from the results.

False data construction probabilities of Targeted LASSO Attacks (TLA), Strategic LASSO Attacks (SLA), Targeted Selective Attacks (TSA) and Strategic Selective Attacks (SSA) are compared in the following. 

\begin{figure}[t!]
\centering
\begin{tabular}{cc}
\subfloat[$Pr(\mathbf{\hat{a}'}_i, \mathbf{a'}_i)$ for TLA]{\includegraphics[width=1.70in, height=1.5in]{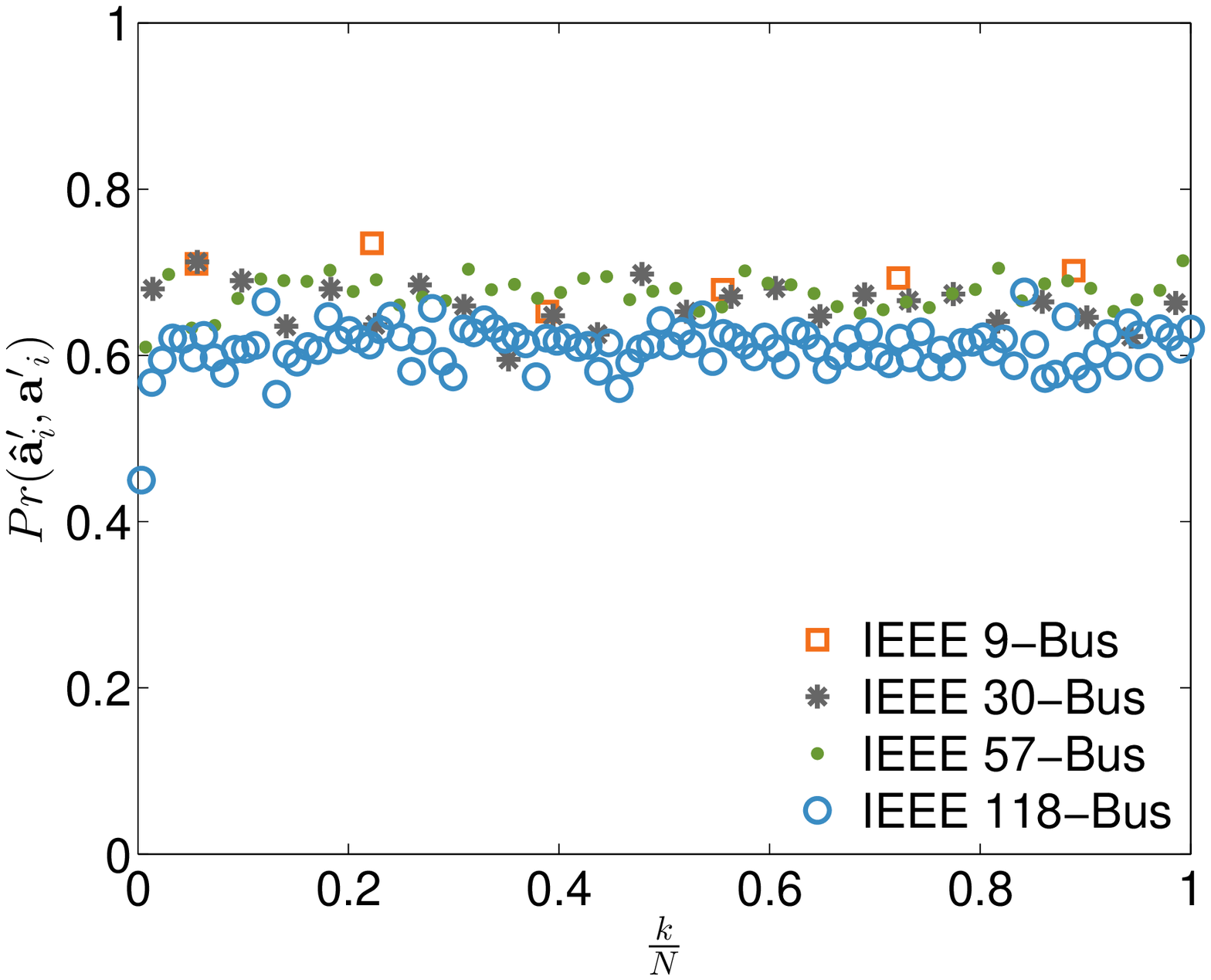}} 
\subfloat[$Pr(\mathbf{\hat{a}}_i, \mathbf{a}_i)$ for TLA]{\includegraphics[width=1.70in, height=1.5in]{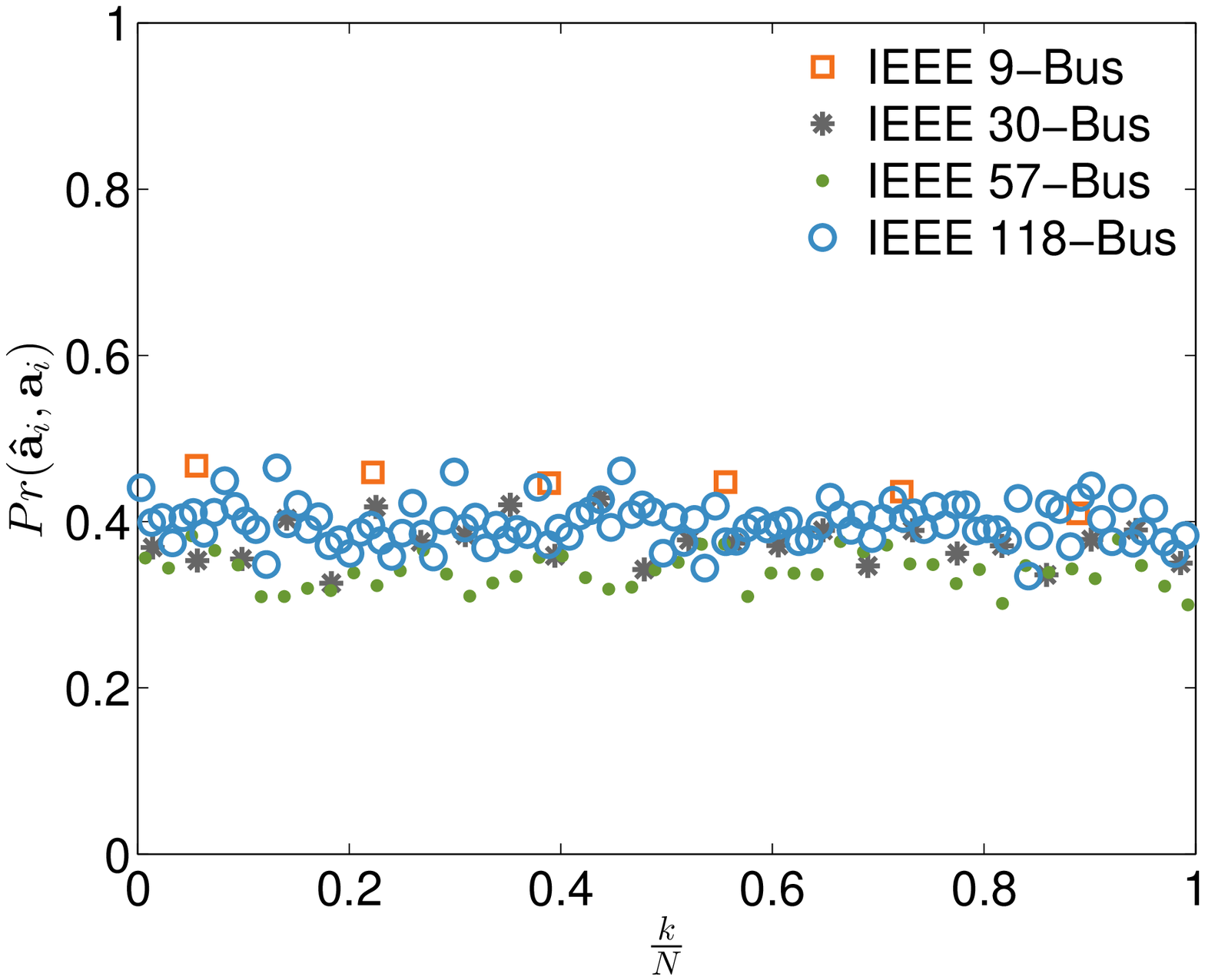}} \\
\subfloat[$Pr(\mathbf{\hat{a}'}_i, \mathbf{a'}_i)$ for TSA]{\includegraphics[width=1.70in, height=1.5in]{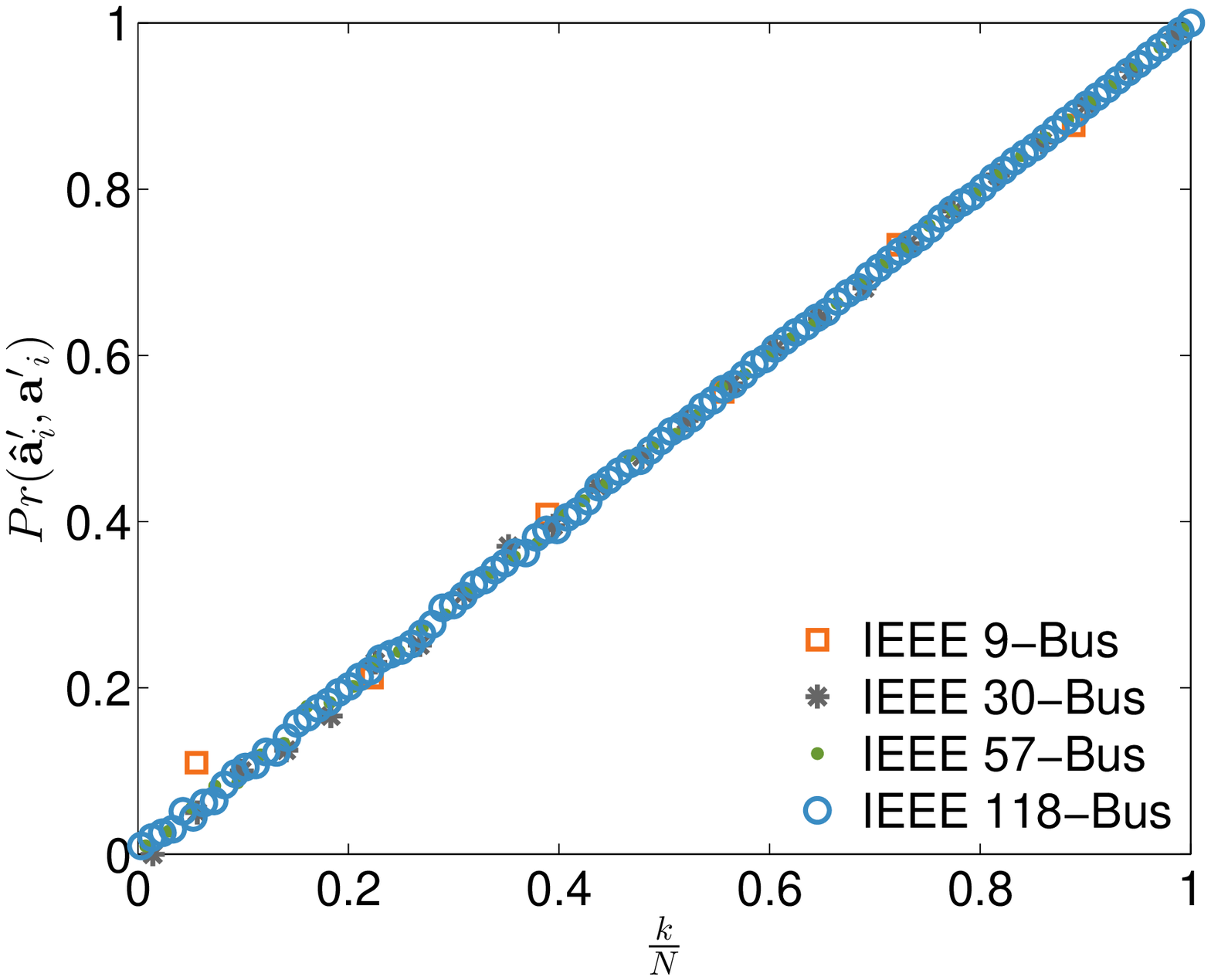}} 
\subfloat[$Pr(\mathbf{\hat{a}}_i, \mathbf{a}_i)$ for TSA]{\includegraphics[width=1.70in, height=1.5in]{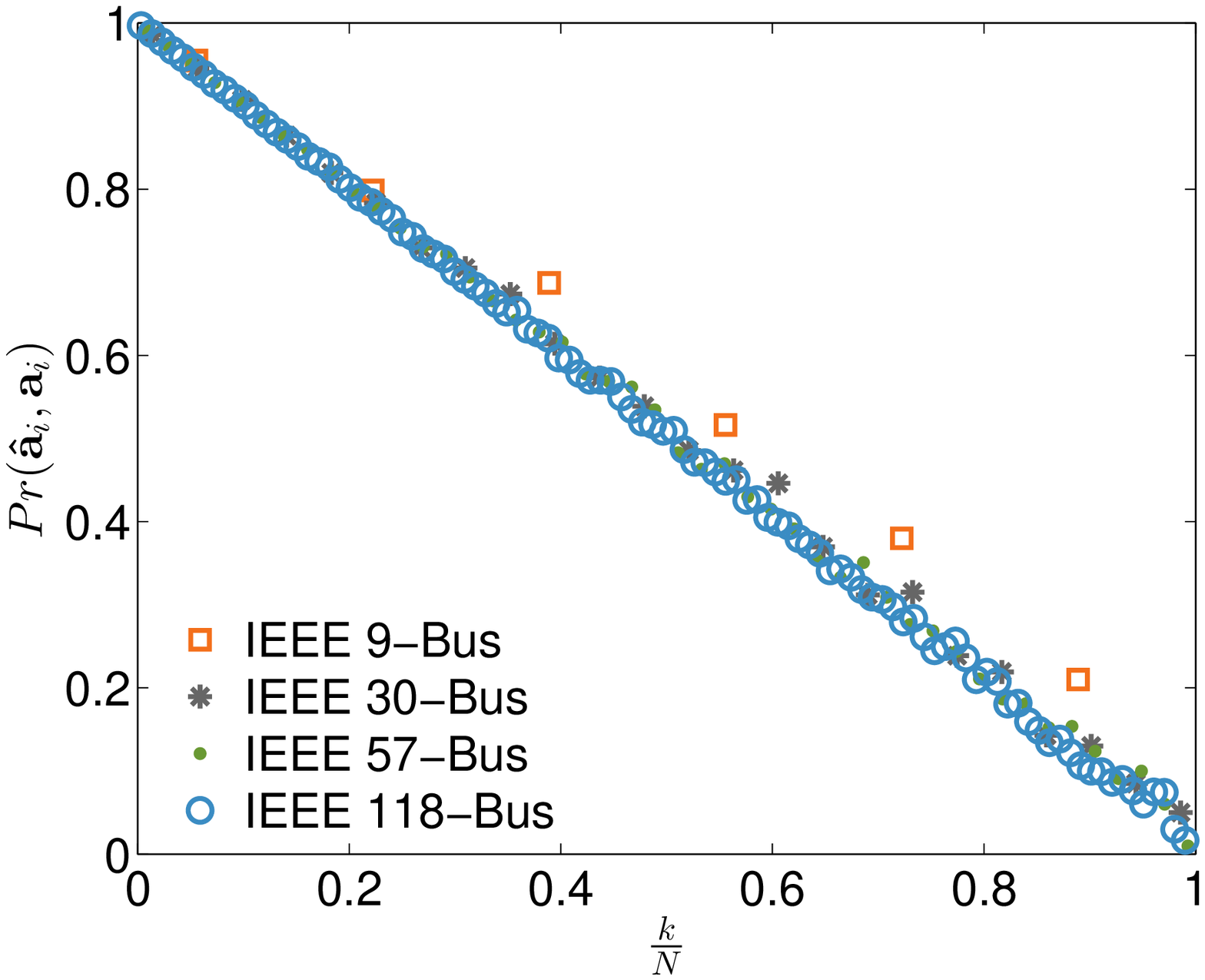}} \\
\end{tabular}
\caption{False data vector construction probabilities for for TLA and TSA. }
\label{fig:c2}
\end{figure}

In Figure \ref{fig:c2}, the experiments for TLA and TSA are analyzed and the changes of false data vector construction probabilities are depicted for a varying number of attack variables, $\frac{k}{N}$, for each test system. The construction probabilities do not increase or decrease smoothly for TLA (Figures \ref{fig:c2}.a and \ref{fig:c2}.b.), since $\lambda$ is computed dynamically by \eqref{eq:lambda} for each realization and test system. Therefore, sparseness is not explicitly controlled in LASSO Attacks. On the other hand, the dynamic computation of $\lambda$ using \eqref{eq:lambda} enables estimation of the sparseness of the attacks and the randomness in $\mathbf{H}$. Therefore, the false data vector $\mathbf{a}$ is constructed with similar probabilities independent of the test system and sparsity level $\frac{k}{N}$ of the attack vectors in the TLA case. For instance, $Pr(\mathbf{\hat{a}'}_i, \mathbf{a'}_i)$ obtains values in the range $[0.5, 0.7]$ in Figure \ref{fig:c2}.a and  $Pr(\mathbf{\hat{a}}_i, \mathbf{a}_i)$ obtains values in the range $[0.3, 0.5]$ in Figure \ref{fig:c2}.b. Since sparseness can be controlled in Selective Attacks, a smooth change of the construction probabilities of false data vector variables is observed for TSA in Figure \ref{fig:c2}.c and Figure \ref{fig:c2}.d.  

\begin{figure}[!h]
\centering
\begin{tabular}{cc}
\subfloat[$Pr(\mathbf{\hat{a}'}_i, \mathbf{a'}_i)$ for SLA]{\includegraphics[width=1.75in, height=1.55in]{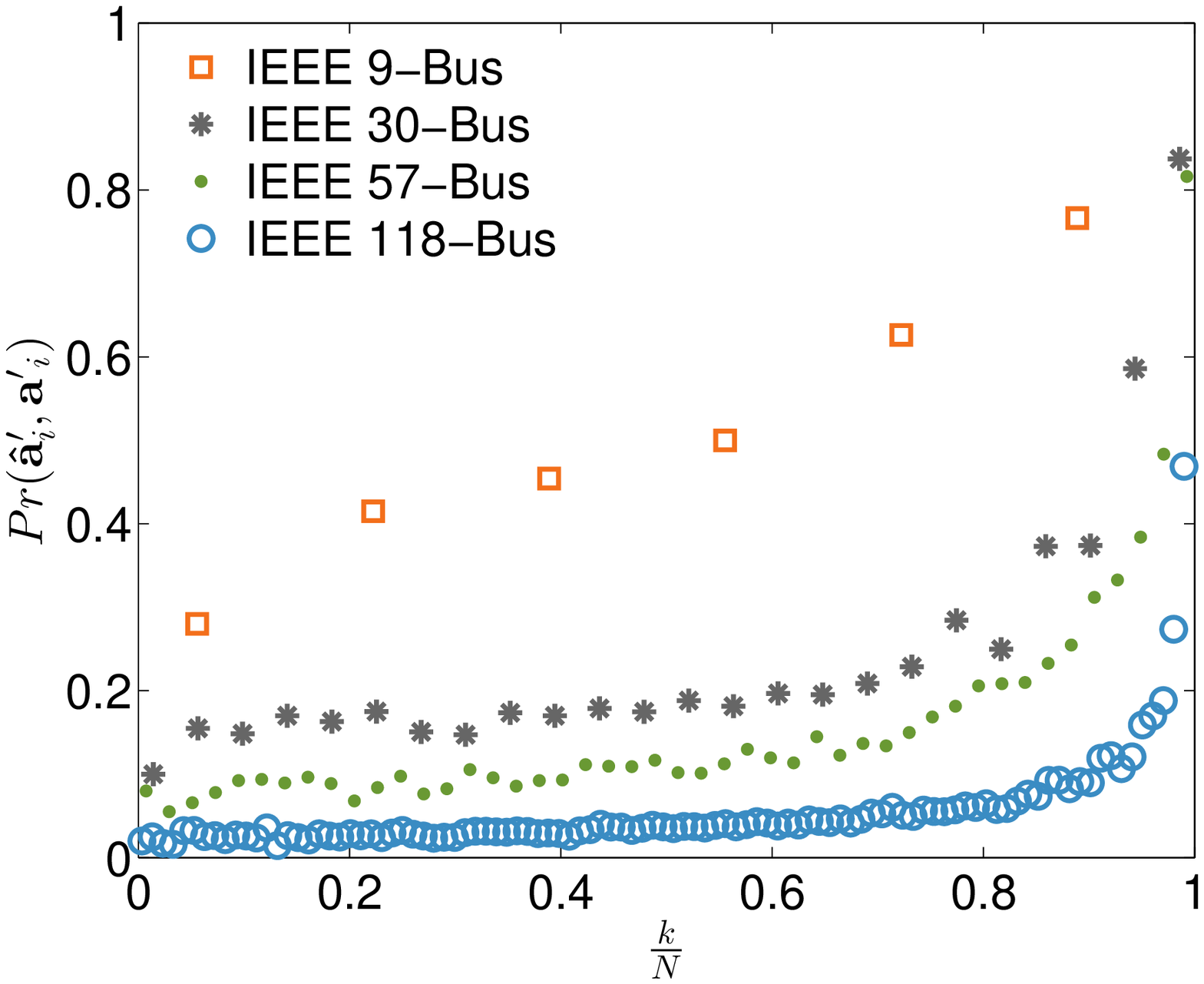}} 
\subfloat[$Pr(\mathbf{\hat{a}}_i, \mathbf{a}_i)$ for SLA]{\includegraphics[width=1.75in, height=1.55in]{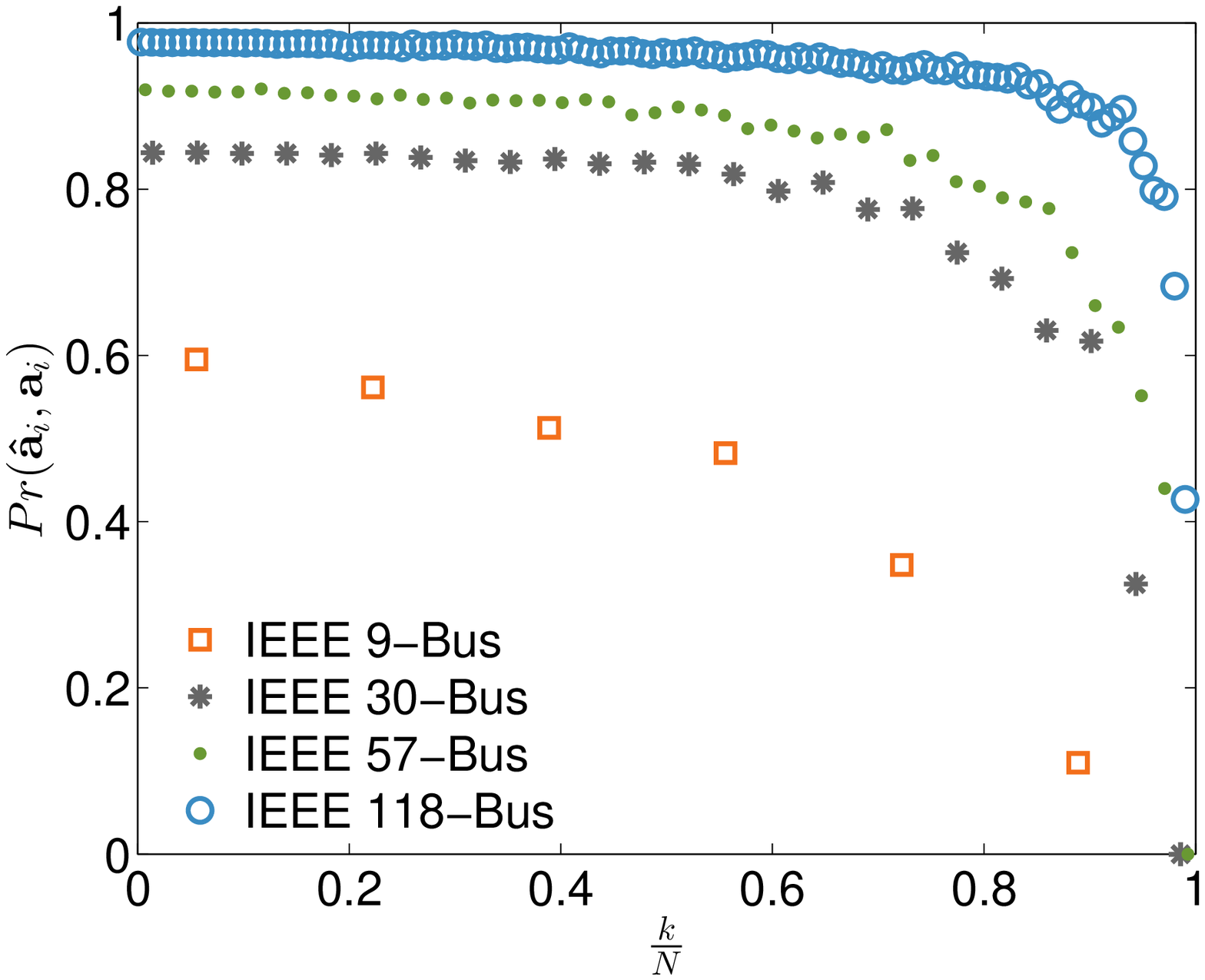}} \\
\subfloat[$Pr(\mathbf{\hat{a}'}_i, \mathbf{a'}_i)$ for SSA]{\includegraphics[width=1.75in, height=1.55in]{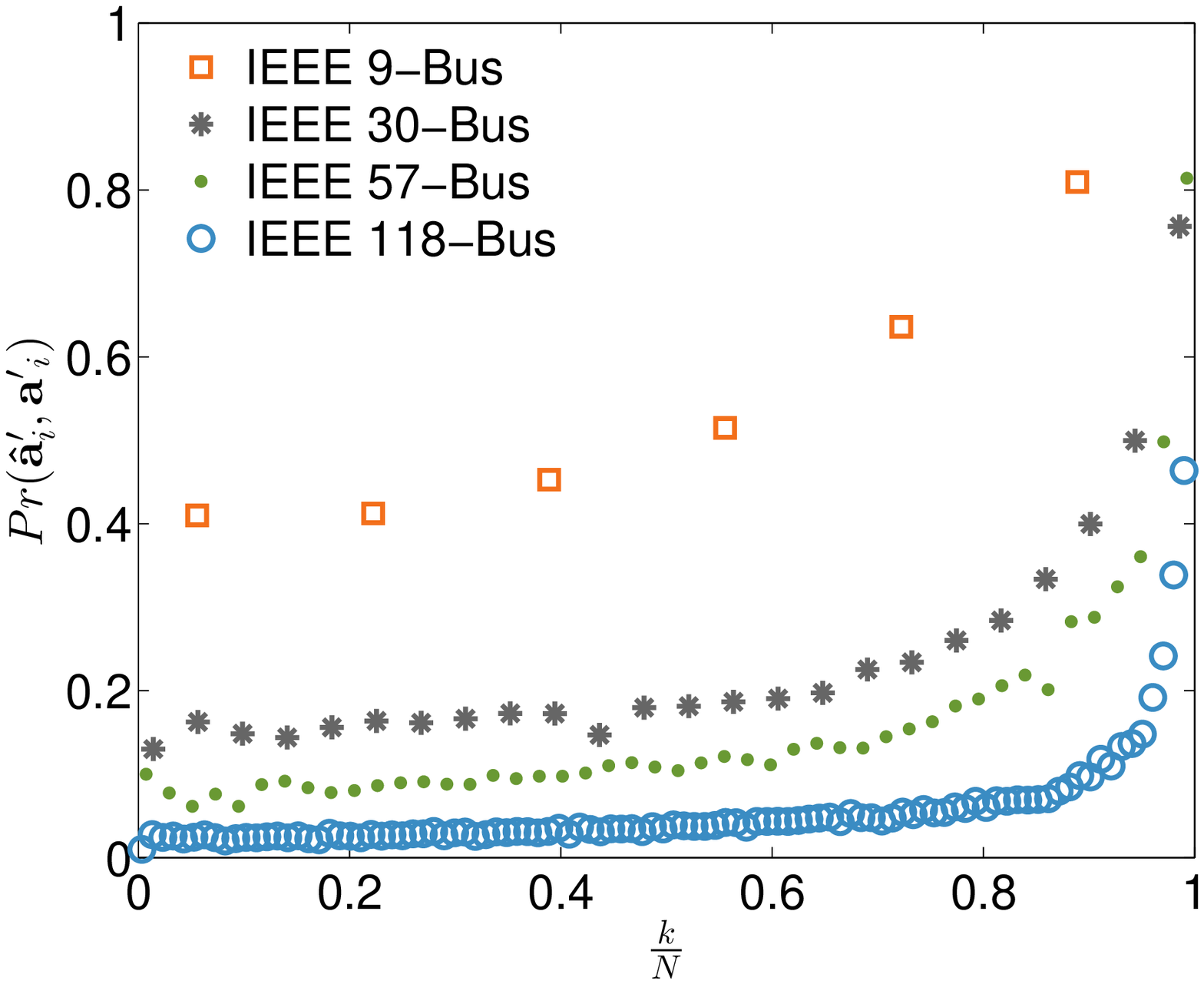}} 
\subfloat[$Pr(\mathbf{\hat{a}}_i, \mathbf{a}_i)$ for SSA]{\includegraphics[width=1.75in, height=1.55in]{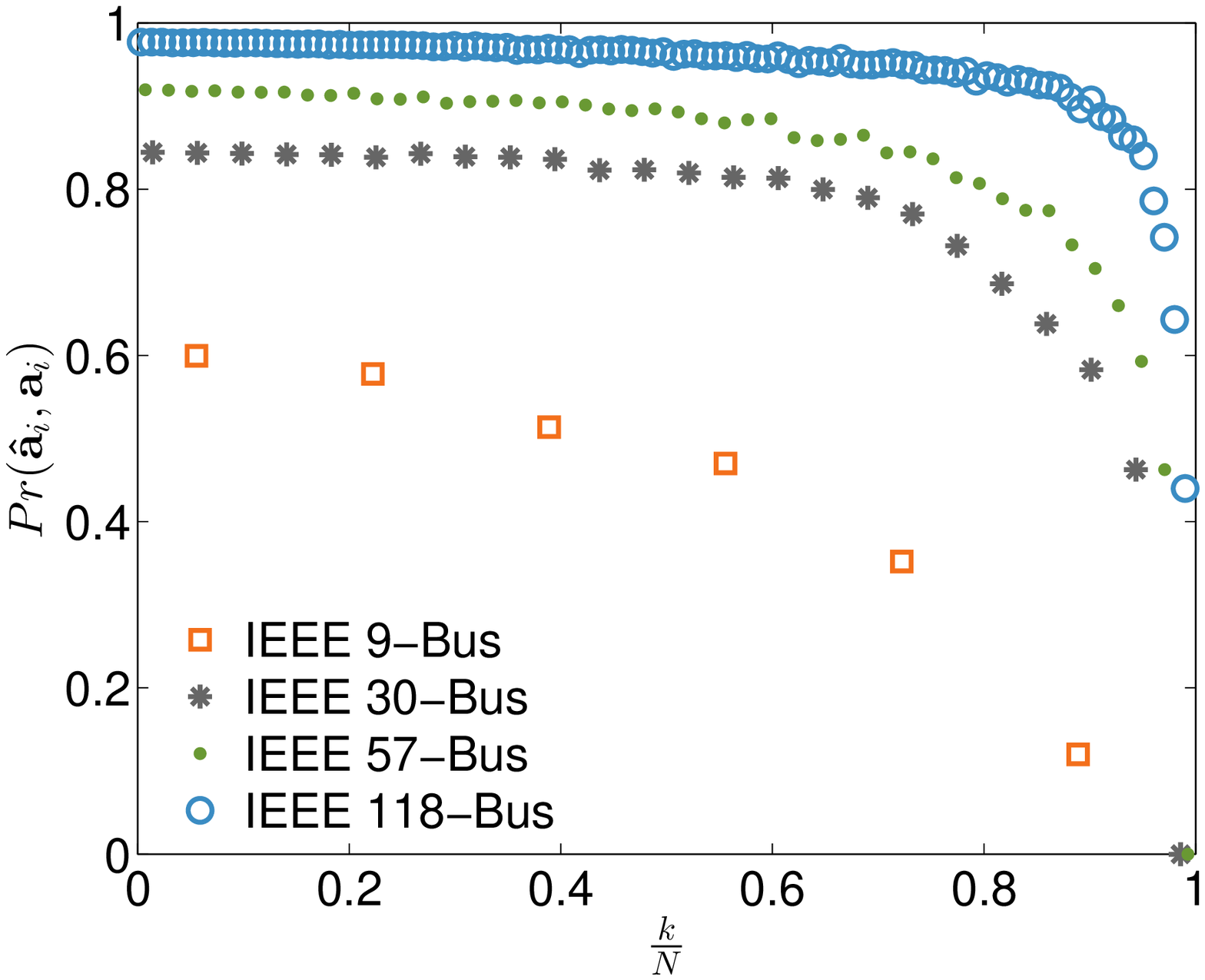}} \\
\end{tabular}
\caption{False data vector construction probabilities for SLA and SSA. }
\label{fig:c3}
\end{figure}

Note that, if the regularization parameter for LASSO is optimized, the solution vectors for LASSO and Regressor selection algorithms coincide \cite{admm}. In Figure \ref{fig:c3}, it can be seen that similar solutions are attainable, \textit{i.e.} both methods construct similar attack vectors. For instance, we observe that the attacked variable construction probabilities increase similarly in Figures \ref{fig:c3}.a and \ref{fig:c3}.c, while secure variable construction probabilities decrease similarly in Figures \ref{fig:c3}.b and \ref{fig:c3}.d for SLA and SSA. 

\subsection{Results for Attack Detection using Distributed Sparse State Vector Estimation}

In this work, our primary interest from the network operator's point of view is the distributed estimation of the state vectors. In this section, we analyze the proposed state vector estimation methods by employing them for the attack detection problem using a modified Normalized Residual Test (NRT) procedure \cite{book}. 

In the attack detection procedure, we first estimate the state vectors $\mathbf{\hat x}$ using the algorithms proposed in Section III. Then, the error of the system is computed as $\| \mathbf{z} - \mathbf{H} \mathbf{\hat x} \|^2 _2$ and the residual of an observed measurement $i$ is given by $\| z_i -\left ( \mathbf{H} \mathbf{\hat x}\right )_{i}\|^2 _2$, where $(\cdot)_i$ denotes the $i$-th element of the argument vector. Following the classical detection criterion, it is declared that the observation $i$ is attacked if $\| z_i - (\mathbf{H} \mathbf{\hat x})_{i} \|^2 _2 > \tau$. Since our goal is to detect the attacks on specific measurements, we do not remove the attacked measurement vectors at each iteration of the algorithm unlike the NRT method proposed in \cite{book}. In addition, such a removal process disturbs the data space. Therefore, the proposed estimation methods should be re-implemented and the regularization parameters should be re-estimated on the updated datasets, leading to additional computational costs. 

In the experiments, both algorithms operate with fixed parameter $C=\frac{1}{2}$ for the regularization parameter $\lambda$. In addition, $\tau$ is chosen as $2\xi_n \| I-\mathbf{H}(\mathbf{H}^T \Sigma_n ^{-1} \mathbf{H})^{-1} \mathbf{H}^T \Sigma_n^{-1}  \| _{\infty}$, where $\xi_n$ and $\Sigma_n$ are the variance and the covariance matrix of the noise $\mathbf{n}$ in \eqref{eq:system_model} respectively, as suggested in \cite{tau}.%

In this section, we construct the attack vectors using \textit{Random False Data Injection Attacks} when the attacker has access to any $k$ meters to construct $k$-sparse attack vectors $\mathbf{a}$, as suggested by Liu, Ning and Reiter \cite{s1}.

\begin{figure*}[ht!]
\centering
\begin{tabular}{cc}
\subfloat[Accuracy for the distributed case.]{\includegraphics[width=2.25in, height=1.75in]{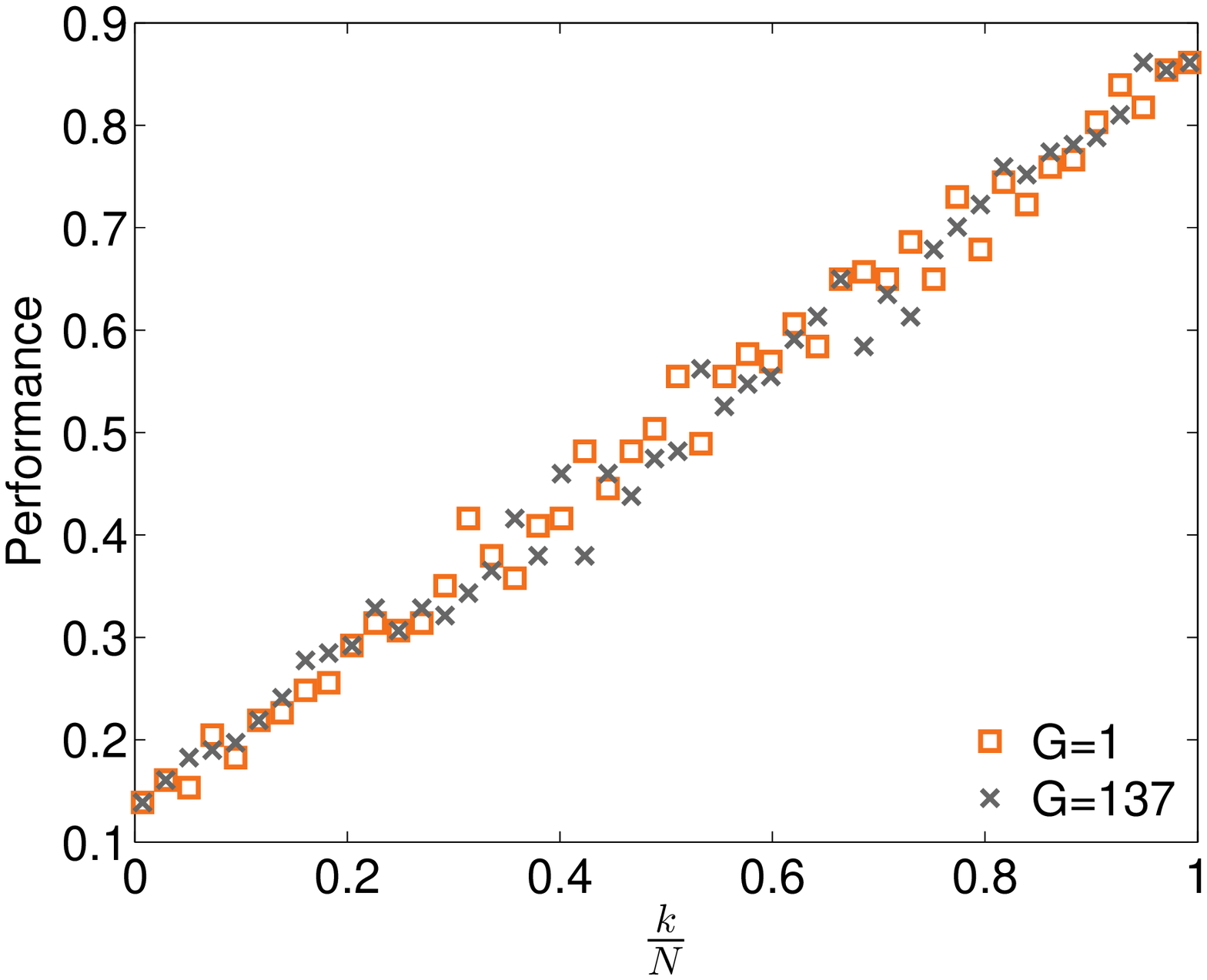}} 
\subfloat[Accuracy for the collective case.]{\includegraphics[width=2.25in, height=1.75in]{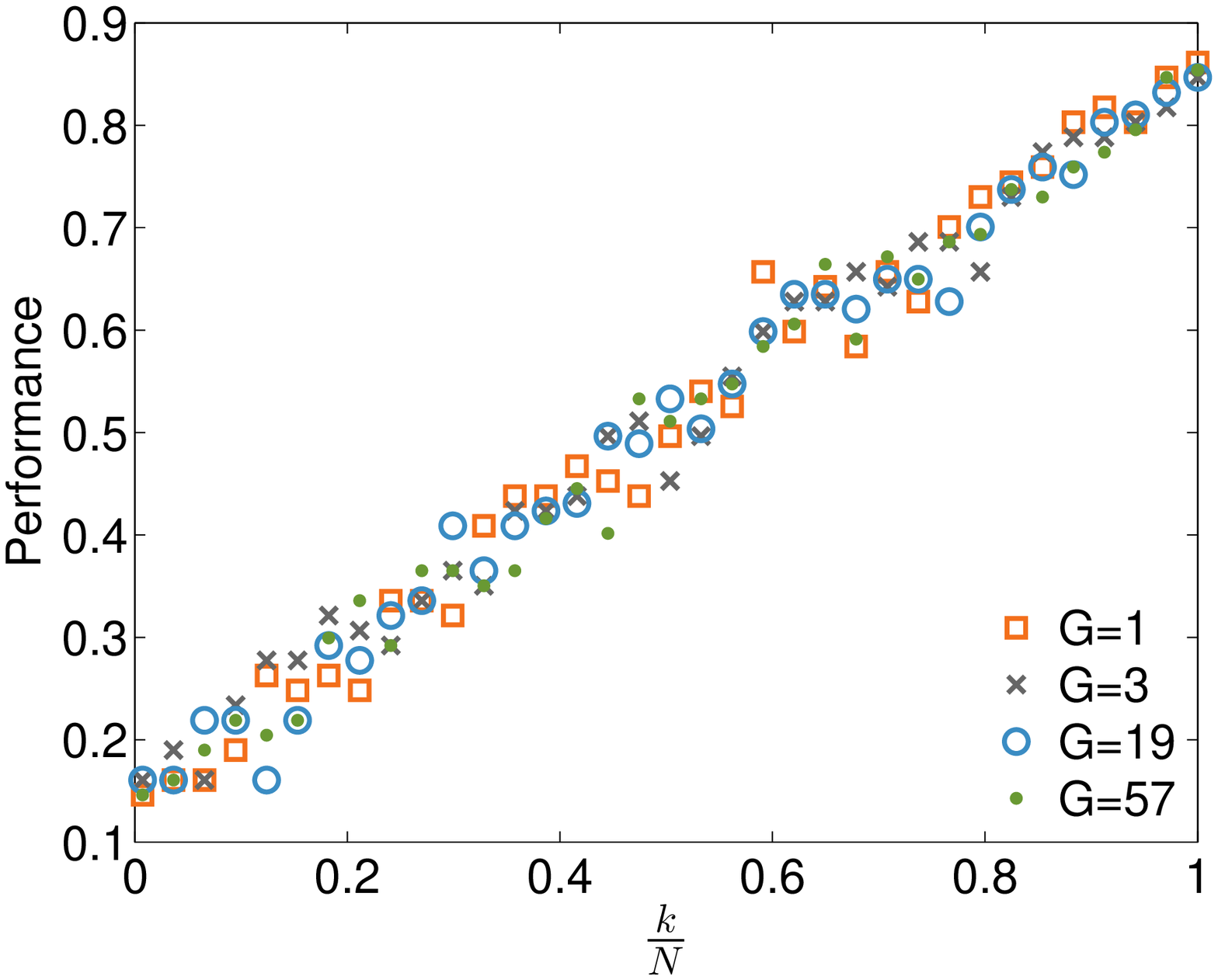}} 
\subfloat[Precision for the distributed case.]{\includegraphics[width=2.25in, height=1.75in]{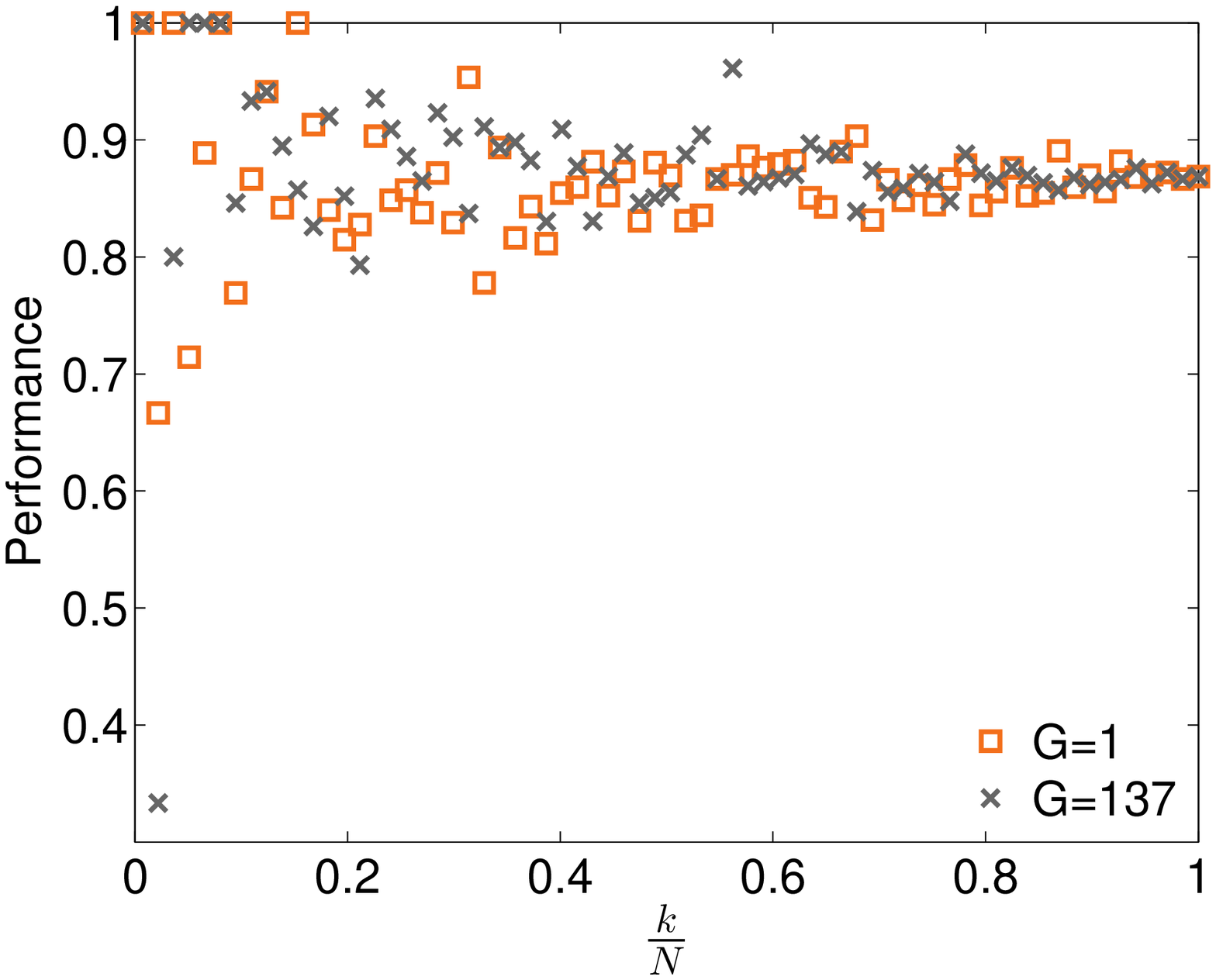}} \\
\subfloat[Precision for the collective case.]{\includegraphics[width=2.25in, height=1.75in]{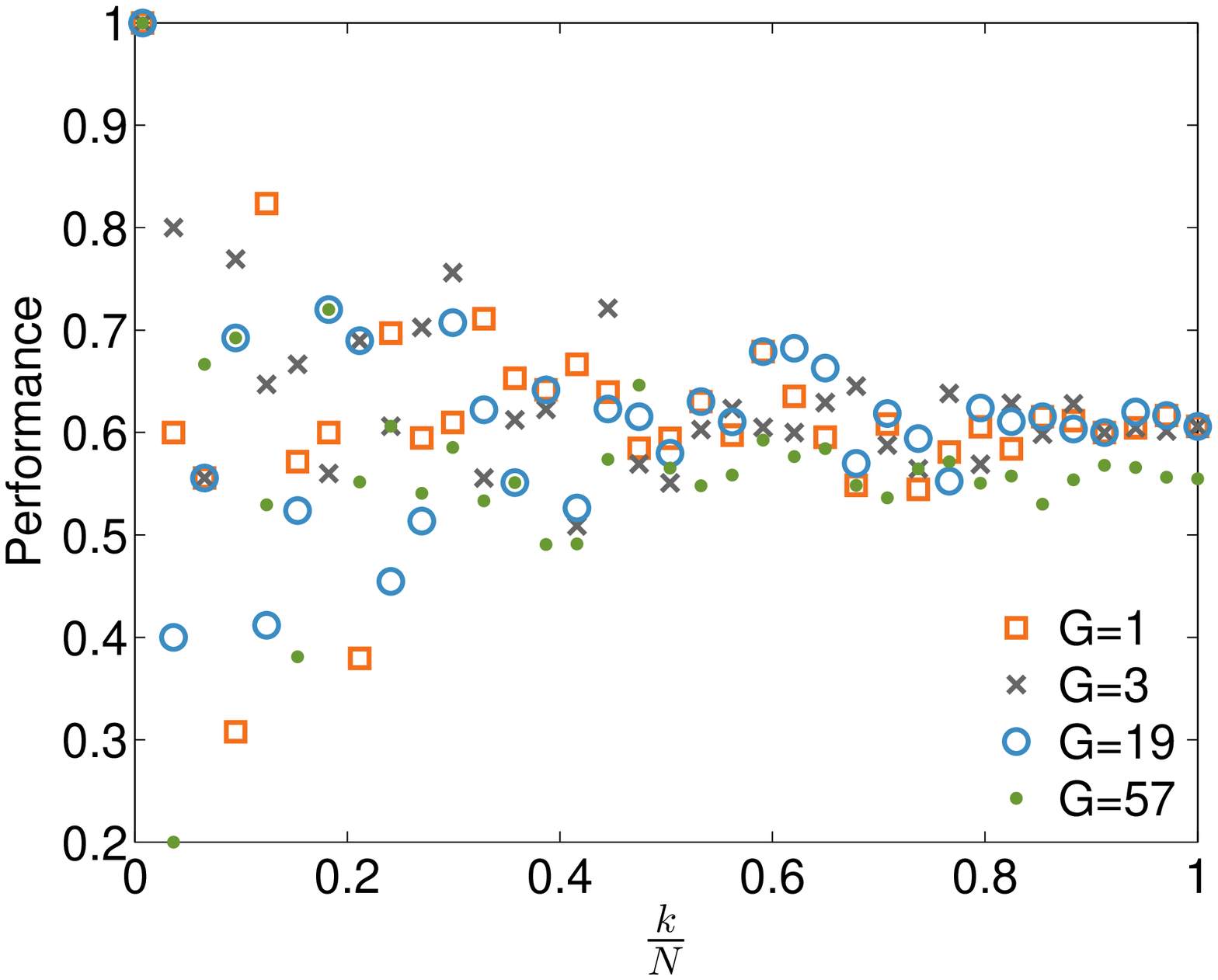}}
\subfloat[Recall for the distributed case.]{\includegraphics[width=2.25in, height=1.75in]{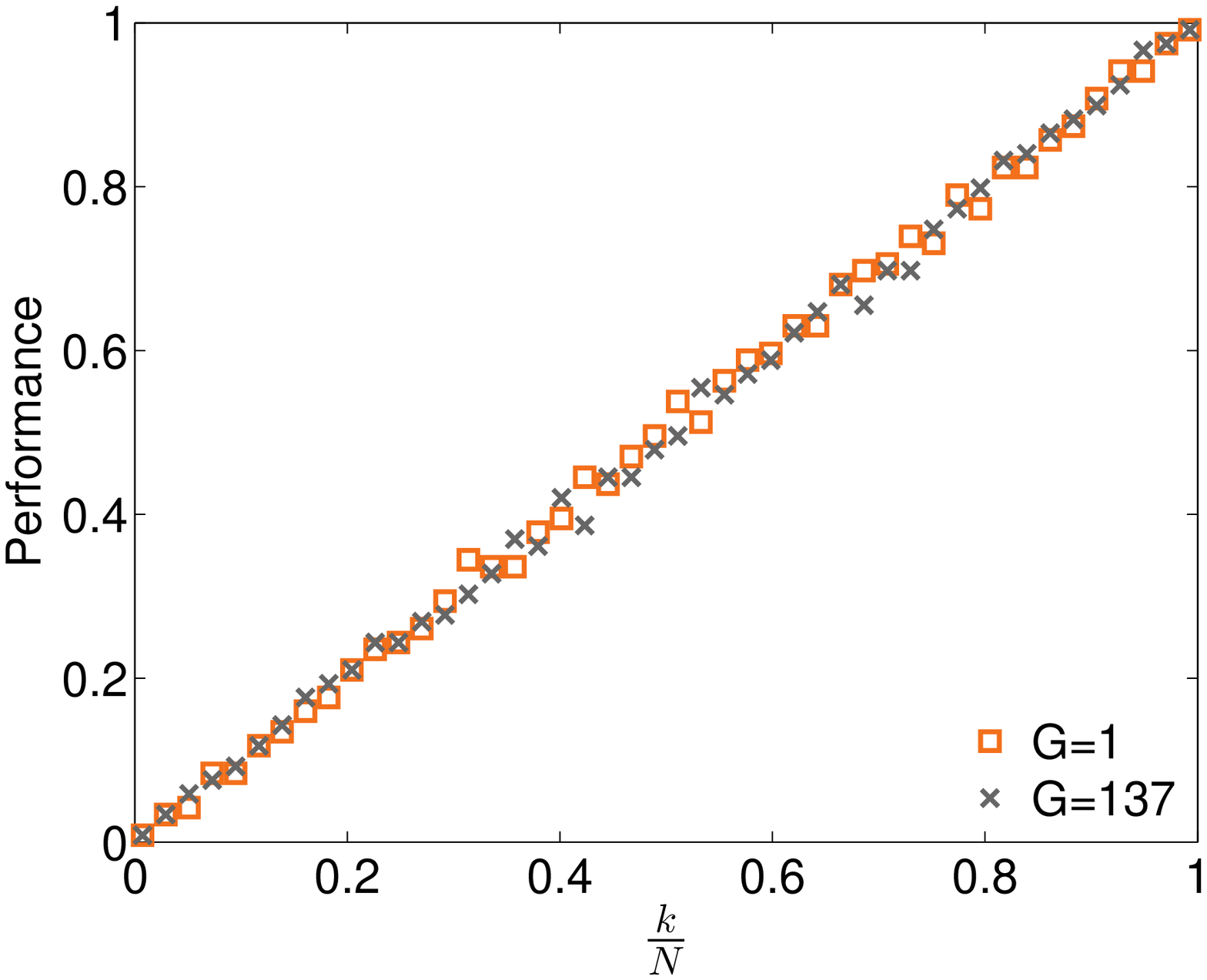}} 
\subfloat[Recall for the collective case.]{\includegraphics[width=2.25in, height=1.75in]{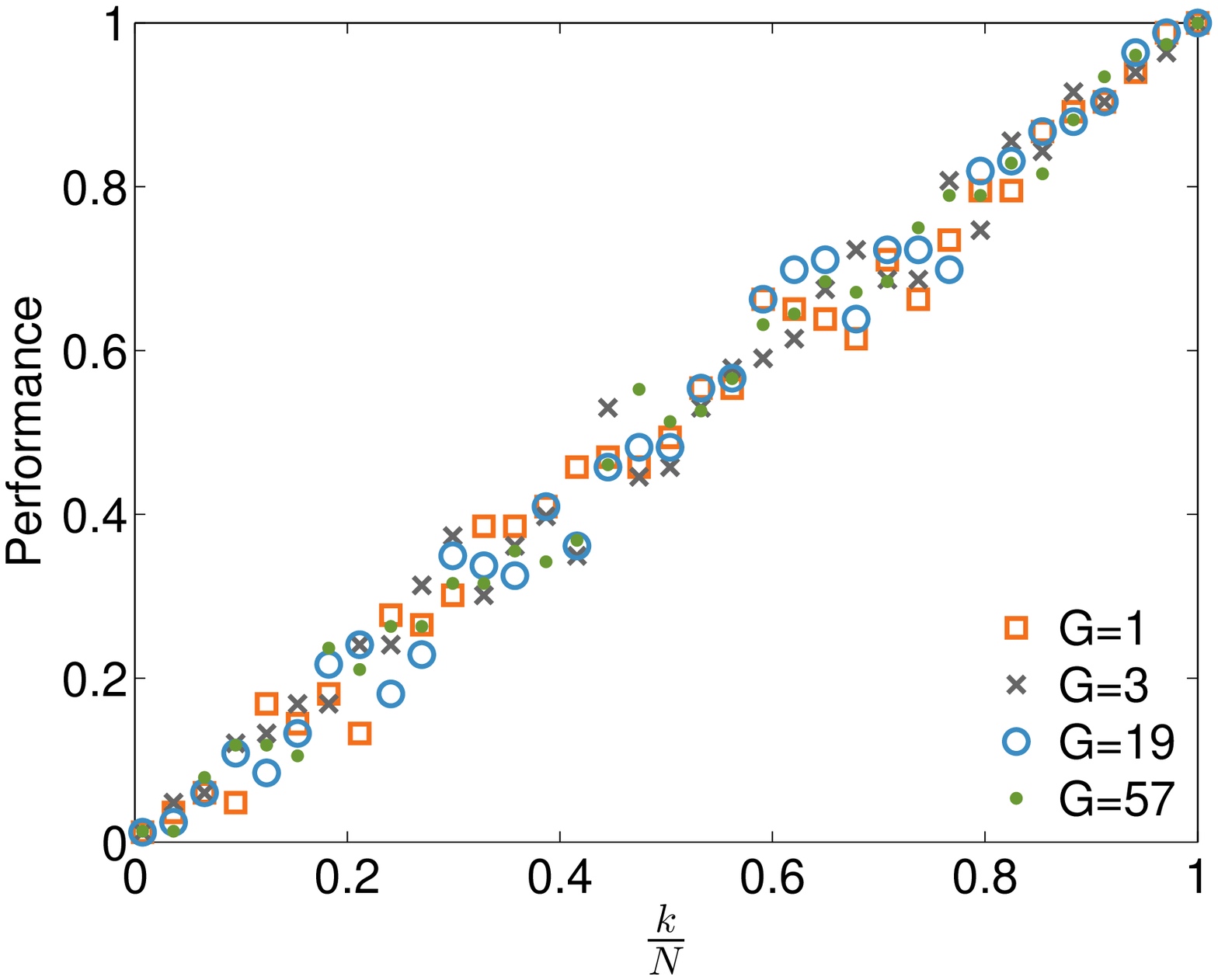}} 
\end{tabular}
\caption{Experiments for the IEEE 57-bus test system with various $G$ values.}
\label{fig:2}
\end{figure*}

Performance indices Precision (\textit{Prec}), Recall (\textit{Rec}) and Accuracy (\textit{Acc}) are defined as
\begin{eqnarray}
\begin{matrix} 
& Prec=\frac{tp}{tp+fp}, & Rec=\frac{tp}{tp+fn} \;, & Acc=\frac{tp+tn}{tp+tn+fn+fp} \;,
\end{matrix} 
\end{eqnarray}
where true positive (\textit{tp}), true negative (\textit{tn}), false positive (\textit{fp}), and false negative (\textit{fn}) are defined in Table~\ref{tab:table2}. For instance, \textit{tp} represents the number of attacked measurements that are correctly detected. On the other hand, \textit{fp} represents the number of secure measurements that are wrongly declared as attacked. Note that, $Prec$ is equal to $Pr(\mathbf{\hat{a}}_i = \mathbf{a}_i)$, which is the probability that a network operator can successfully detect $k$ specific attacks for all $ \mathbf{\hat{a}}_i \neq 0$.

\begin{table}[htbp]
  \centering
  \caption{Definitions of \textit{tp}, \textit{fp}, \textit{tn}, and \textit{fn} }
\begin{tabular}{r|c|c|}
\multicolumn{1}{r}{}
 &  \multicolumn{1}{c}{Attacked}
 & \multicolumn{1}{c}{Secure} \\
\cline{2-3}
Classified as Attacked & \textit{tp} & \textit{fp} \\
\cline{2-3}
Classified as Secure & \textit{fn} & \textit{tn} \\
\cline{2-3}
\end{tabular}
  \label{tab:table2}%
\end{table}%

Results for different numbers of measurement clusters, $G$, are considered in Figure \ref{fig:2}. In this experiment, each operator has access to locally observed measurements, state vectors and submatrices. The simulated cases are $G=|N|$ and $G=|D|$, for distributed and collective state estimation algorithms respectively. Therefore, $G=|N|$ and $G=|D|$ are the extreme cases for distributed processing scenarios. However, the algorithms have similar performance for different values of $G$ in Figure \ref{fig:2}, which shows that the optimality loss with respect to centralized strategies is small in the simulated settings.

\begin{figure}[ht]
\centering
\begin{tabular}{cc}
\subfloat[Distributed detection for 9-bus.]{\includegraphics[width=1.75in, height=1.55in]{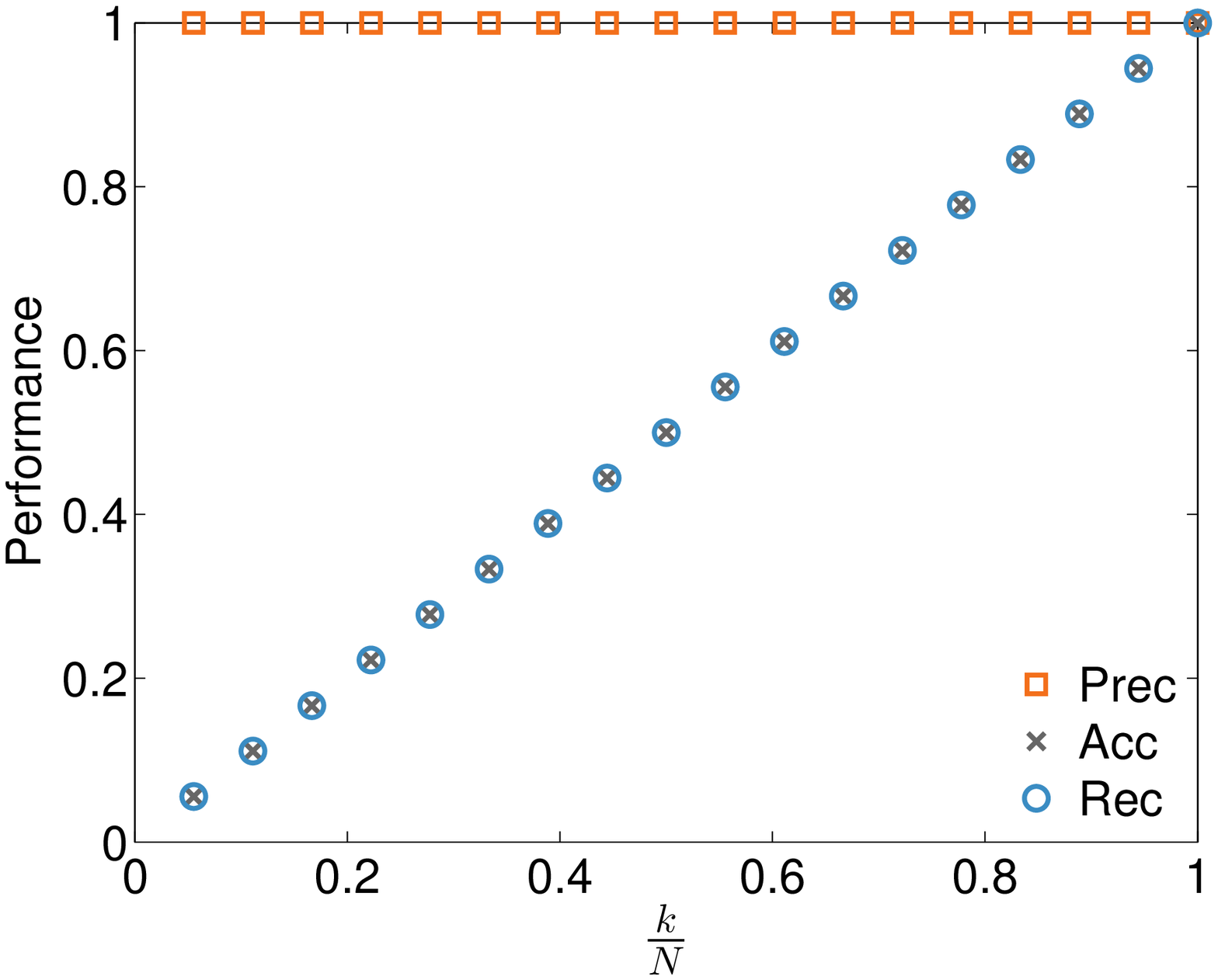}} 
\subfloat[Collective detection for 9-bus.]{\includegraphics[width=1.75in, height=1.55in]{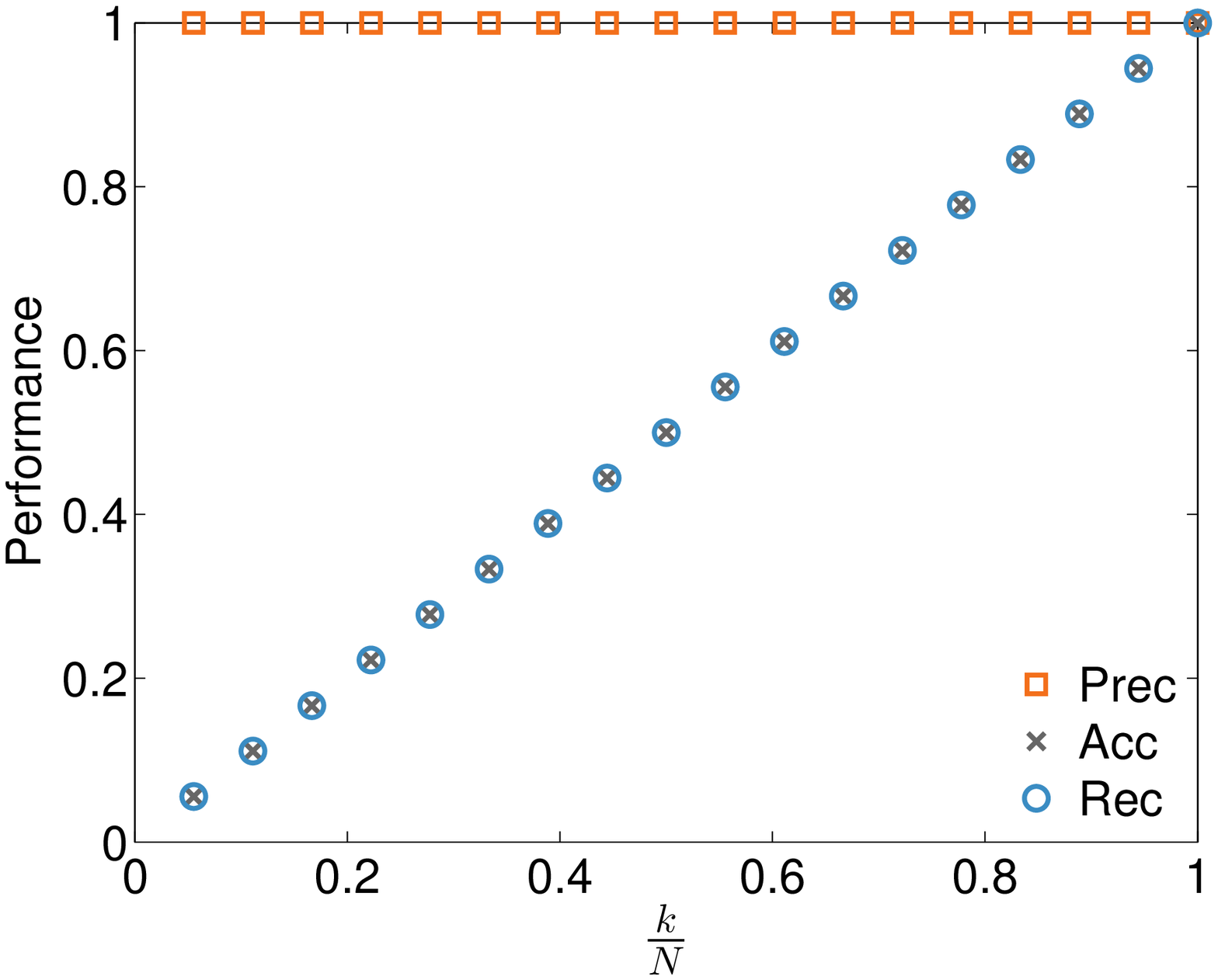}} \\
\subfloat[Distributed detection for 57-bus.]{\includegraphics[width=1.75in, height=1.55in]{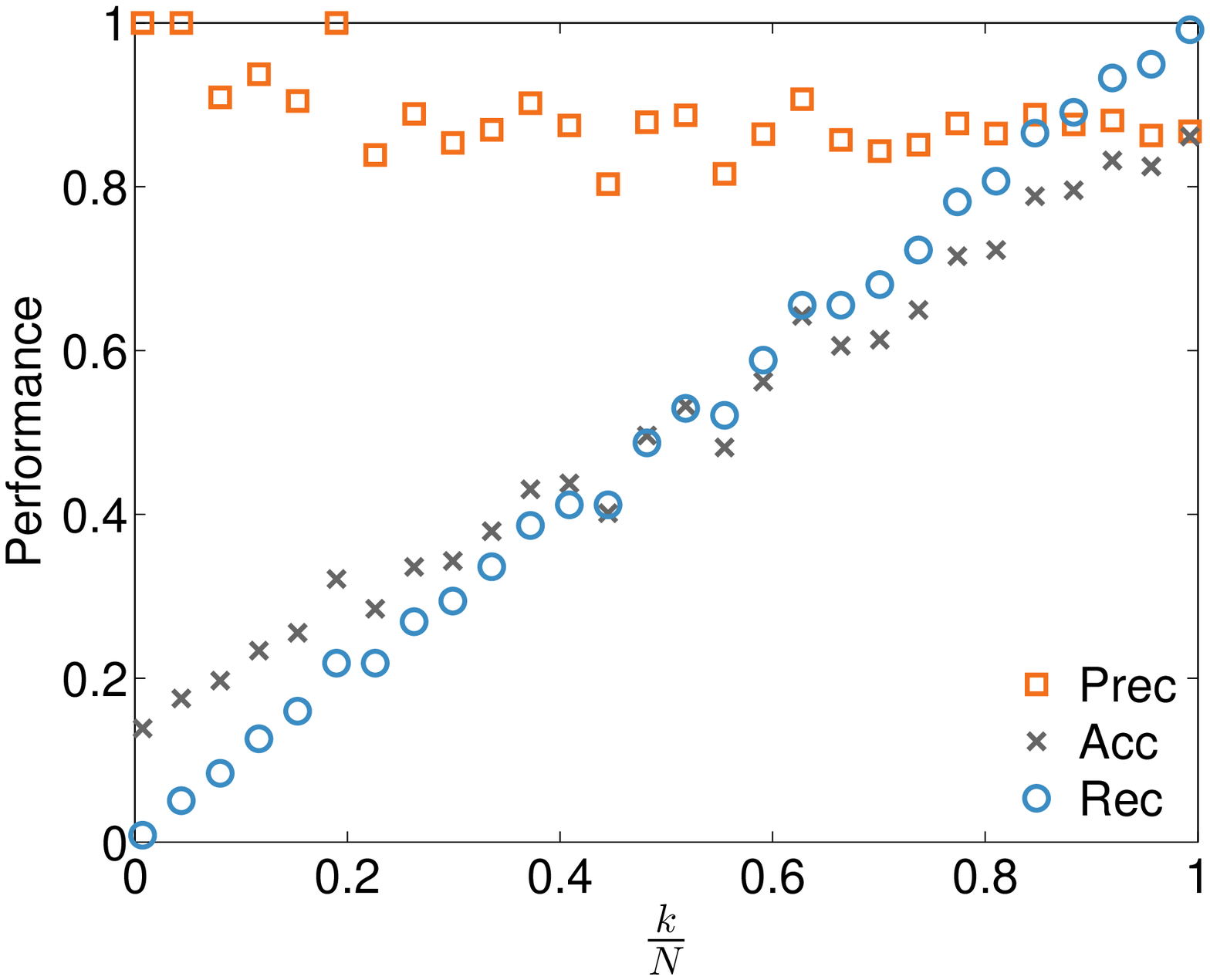}} 
\subfloat[Collective detection for 57-bus.]{\includegraphics[width=1.75in, height=1.55in]{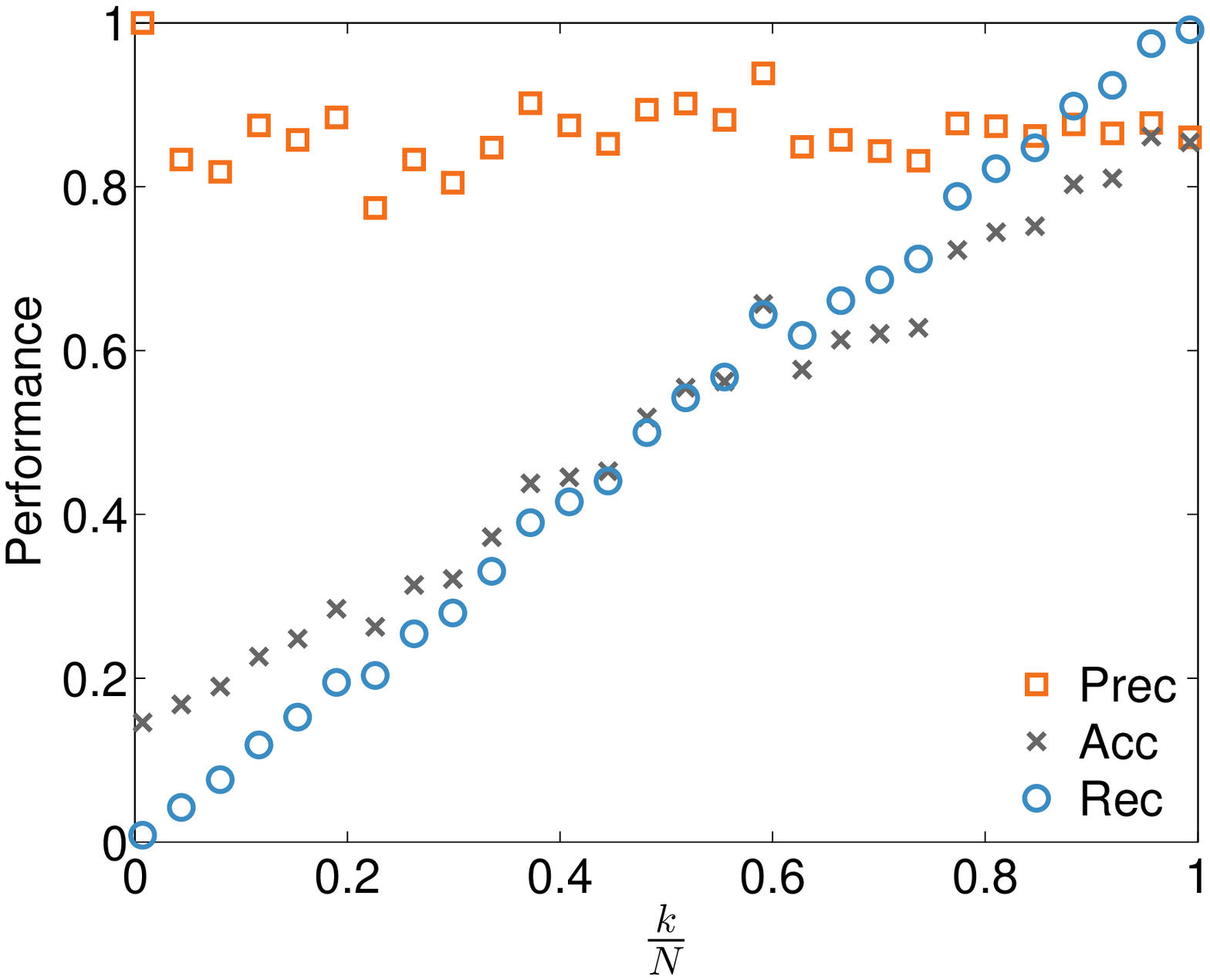}} 
\end{tabular}
\caption{Distributed estimation performance indices for IEEE 9-bus and IEEE 57-bus test systems.}
\label{fig:1}
\end{figure}

Figure \ref{fig:1} shows the results of distributed and collective state vector estimation algorithms for $G=1$. Note that, the case $G=1$ represents a centralized processing scenario in which all of the observed measurements and the whole Jacobian measurement matrix are available to network operators. It can be seen that the performance values of distributed and collective estimation methods are similar for the IEEE-9-bus test system in Figure \ref{fig:1}.a and Figure \ref{fig:1}.b respectively. However, the results in Figures \ref{fig:1}.c and \ref{fig:1}.d show that for low values of $k/N$ the precision fluctuates and stabilizes around 0.9 as $k/N$ increases for the IEEE-57-bus test system. In addition, the slopes of the curves representing the increases of accuracy and recall values are slightly smaller in Figures \ref{fig:1}.c and \ref{fig:1}.d than the ones for the IEEE-9-bus test system case.

\subsection{Results for Distributed and Collective Sparse Attacks}


In order to measure the detectability of the attacks from the perspective of the network operators, $Error= \| z_i - (\mathbf{H} \mathbf{\hat x})_i\|^2 _2 $ is considered. Throughout this section, both algorithms operate with fixed parameter $C=\frac{1}{2}$.


\begin{figure}[t]
\centering
\begin{tabular}{cc}
\subfloat[$Pr(\mathbf{\hat{a}'}_i, \mathbf{a'}_i)$ of distributed attacks.]{\includegraphics[width=1.75in, height=1.55in]{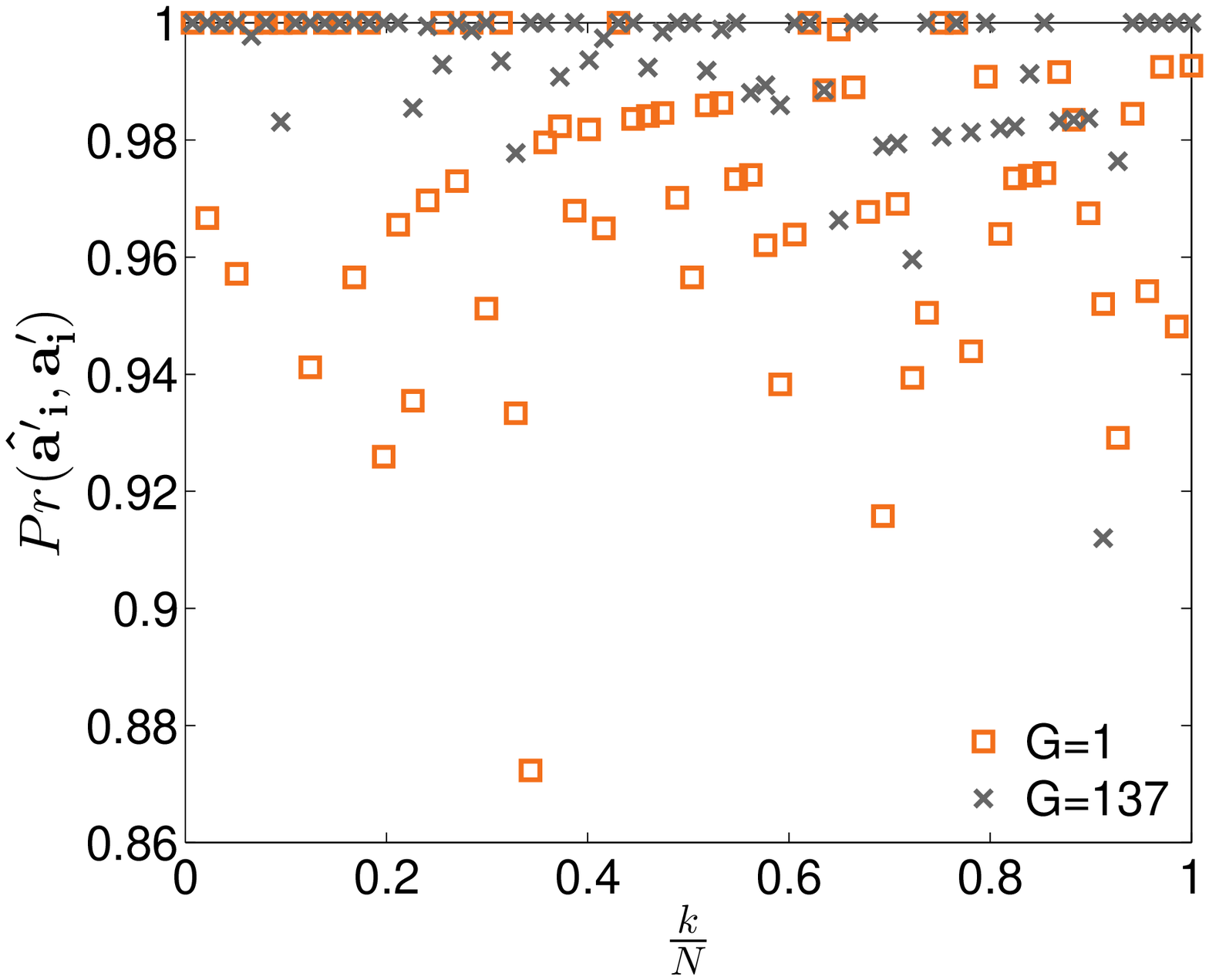}} 
\subfloat[Error of distributed attacks.]{\includegraphics[width=1.75in, height=1.55in]{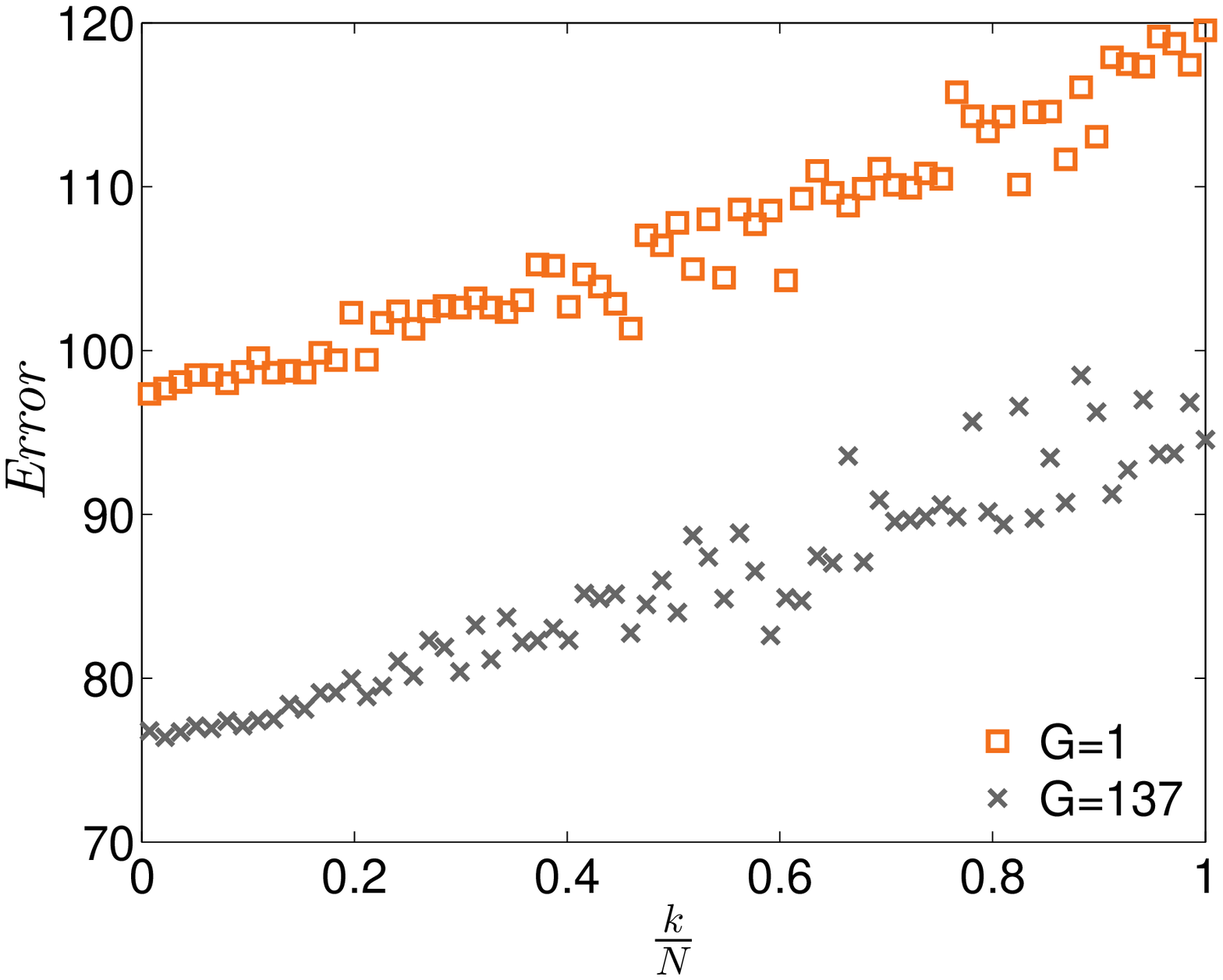}} \\
\subfloat[$Pr(\mathbf{\hat{a}'}_i, \mathbf{a'}_i)$ of collective attacks.]{\includegraphics[width=1.75in, height=1.55in]{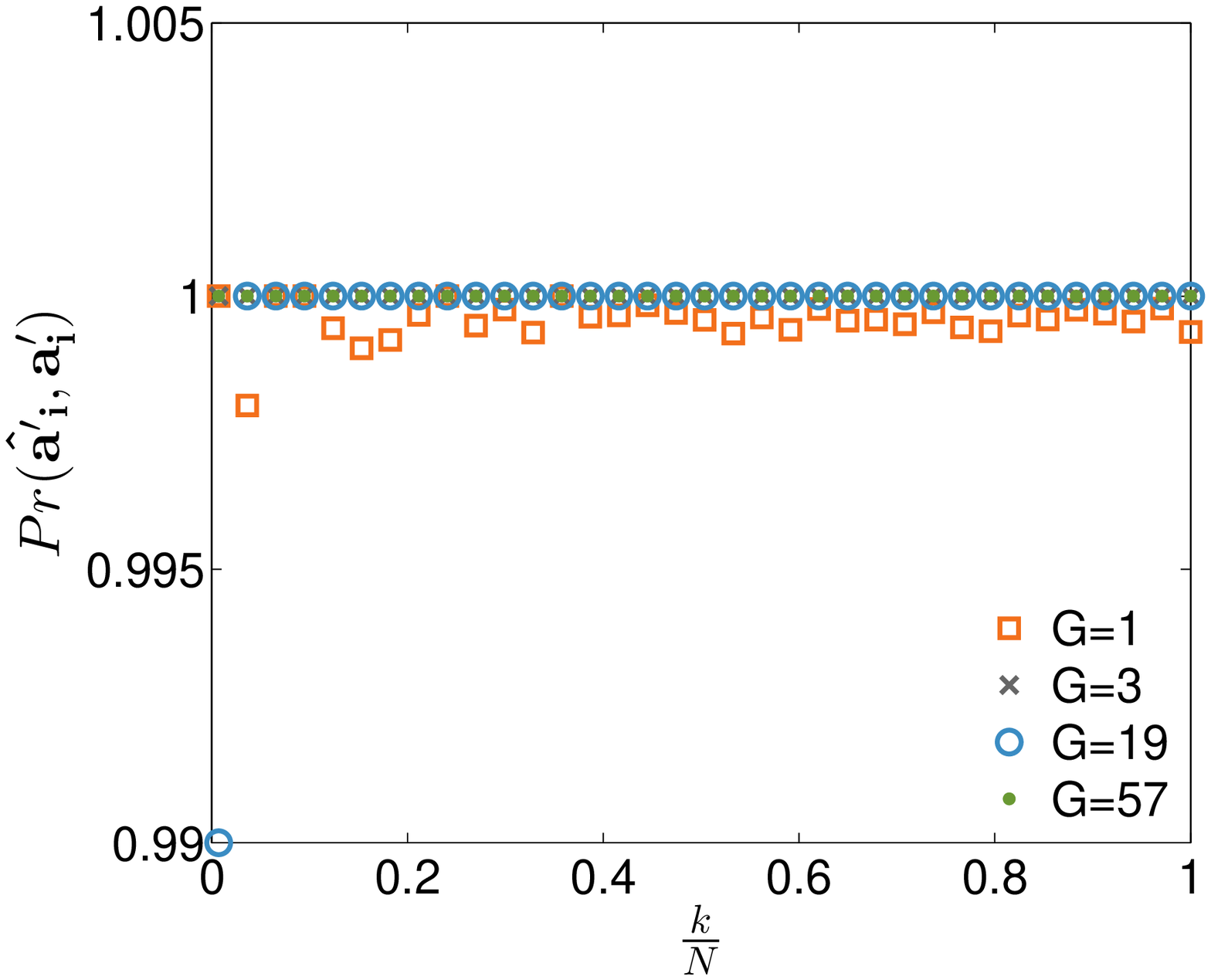}} 
\subfloat[Error of collective attacks.]{\includegraphics[width=1.75in, height=1.55in]{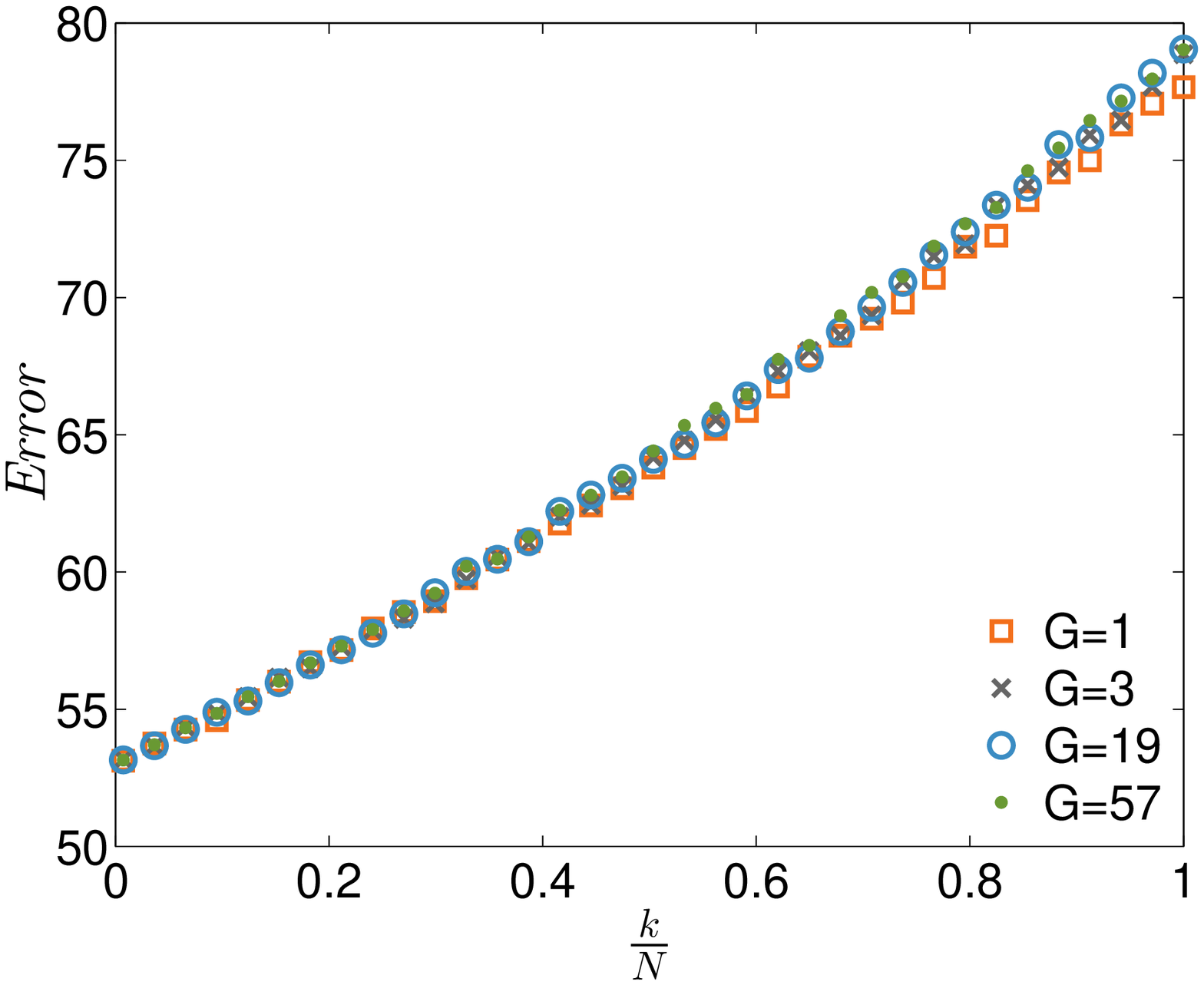}} 
\end{tabular}
\caption{Experiments for the IEEE 57-bus test system.}
\label{fig:3}
\end{figure} 

In Figure \ref{fig:3}, the results of the distributed attack experiments for the IEEE-57-bus test system are shown. Remarkably, the proposed algorithms are capable of successfully injecting data with high probability for a large range of sparsity ranges. However, it can be seen in Figures \ref{fig:3}.a and \ref{fig:3}.b that $Pr(\mathbf{\hat{a}'}_i, \mathbf{a'}_i)$ decreases and $Error$ increases as $G$ decreases. This is due to the fact that the optimization is very sensitive to the optimization of the regularization parameter $\lambda$. Interestingly, in the simulation settings evaluated for this paper, it has been observed that the proposed algorithms are more robust to variations of $\lambda$ when smaller values of $G$ are considered. It is known \cite{Sundeep} that  the optimization of the regularization parameter in the centralized case is hard. Surprisingly, as the fragmentation of the optimization problem increases, i.e., for lower values of $G$, the performance of the algorithm is less sensitive to the tuning of the regularization parameter. For instance, the injection vectors are computed and the optimization variables are updated locally in each group with respect to a global regularization parameter in distributed and collective attacks. In other words, the group-wise local regularization paths (i.e., the set of solutions) are computed and used to approximate a global regularization path. Since the paths of Group LASSO are piecewise differentiable, approximating the global path by the local paths may be a challenge as $G$ increases. A solution to this challenge is to compute group-wise parameters in an adaptive scheme \cite{bach_lasso}. 

\section{Conclusion}

In this paper, we have considered centralized and distributed models for sparse attack construction and state estimation in the smart grid. For a centralized scenario, two methods, LASSO Attacks and Selective Attacks, have been introduced for the construction of false data vectors and attack vectors for a given attack model. The presented methods are used in two well-known attack models, Targeted and Strategic Attacks. 

We have shown that Selective Attacks provide control of the sparsity of the attack vectors, explicitly. Therefore, a construction method has been proposed for false data and attack vectors, which contain a given number of attacked and secure variables. Incidentally, the randomness of the parameters of the attack models may decrease the unobservability of the attack vectors and the control for the construction of false data vectors. For instance, random construction of the sub-matrices in Targeted Attacks may inject additional randomness into the probabilities of constructing false data vectors $\mathbf{a}$.

For the case in which the distributed nature of the network is considered, new distributed sparse state vector estimation and attack detection methods have been introduced. In the \textit{Distributed State Vector Estimation} method, it is assumed that the observed measurements are distributed in clusters in the network. The state vectors are estimated using local data measurements in the clusters by either local network operators or PMUs. The estimates are then updated by centralized processors. In \textit{Collaborative Sparse State Vector Estimation}, operators estimate a subset of variables of the state vectors. Therefore, state vector variables are assumed to be distributed in groups and accessed by the network operators locally. In this scenario, network operators compute their local estimates and send the estimated values to a centralized network operator in order to update the estimated values.

In the experiments, it has been observed that both state vector estimation methods perform similarly for a varying number of attacks in different test systems. Besides, accuracy and precision values of the proposed methods decrease as the system size increases and the performance values do not change as the number of clusters increases. In other words, we can achieve similar performance when we implement the algorithms in centralized ($G=1$) and massively distributed scenarios ($G=N$ or $G=D$). 

When the \textit{Distributed Sparse Attacks} model is considered, it is assumed that attackers process only local measurements in order to achieve a consensus for attack vectors. In the \textit{Collective Sparse Attacks} case, the topological information of the network and the measurements is available to attackers. However, the attackers employ attacks on variable groups of state vectors. 

It has been observed in the experiments that the \textit{Collective Sparse Attacks} model performs better than the \textit{Distributed Sparse Attacks} model for the construction of unobservable attack vectors. Surprisingly, better performance of the algorithm with higher $G$ values is achieved than smaller $G$ values when larger systems are considered. This is due to the fact that one of the challenges of the proposed methods is the estimation of algorithm parameters, e.g., the maximum number of iterations and the regularization parameter $\lambda$. For the case in which the sparsity degree, $k$, of the solution vectors are known \textit{a priori}, regressor selection algorithms can be employed in order to control the sparsity of the solutions.   

\bibliographystyle{IEEEtran}
\bibliography{IEEEabrv,jsac_13_proof_bibliography}

\begin{thebibliography}{10}
\providecommand{\url}[1]{#1}
\csname url@samestyle\endcsname
\providecommand{\newblock}{\relax}
\providecommand{\bibinfo}[2]{#2}
\providecommand{\BIBentrySTDinterwordspacing}{\spaceskip=0pt\relax}
\providecommand{\BIBentryALTinterwordstretchfactor}{4}
\providecommand{\BIBentryALTinterwordspacing}{\spaceskip=\fontdimen2\font plus
\BIBentryALTinterwordstretchfactor\fontdimen3\font minus
  \fontdimen4\font\relax}
\providecommand{\BIBforeignlanguage}[2]{{%
\expandafter\ifx\csname l@#1\endcsname\relax
\typeout{** WARNING: IEEEtran.bst: No hyphenation pattern has been}%
\typeout{** loaded for the language `#1'. Using the pattern for}%
\typeout{** the default language instead.}%
\else
\language=\csname l@#1\endcsname
\fi
#2}}
\providecommand{\BIBdecl}{\relax}
\BIBdecl

\bibitem{s1}
Y.~Liu, P.~Ning, and M.~K. Reiter, ``False data injection attacks against state
  estimation in electric power grids,'' in \emph{Proc. 16th ACM Conf. Comput.
  Commun. Security}, ser. CCS '09.\hskip 1em plus 0.5em minus 0.4em\relax New
  York, NY, USA: ACM, 2009, pp. 21--32.

\bibitem{lalitha}
L.~Sankar, S.~Kar, R.~Tandon, and H.~V. Poor, ``Competitive privacy in the
  smart grid: An information-theoretic approach,'' in \emph{Proc. 2nd IEEE Int.
  Conf. Smart Grid Communications}, Brussels, Belgium, Oct. 2011, pp. 220--225.

\bibitem{cn}
E.~Cotilla-Sanchez, P.~Hines, C.~Barrows, and S.~Blumsack, ``Comparing the
  topological and electrical structure of the {N}orth {A}merican electric power
  infrastructure,'' \emph{IEEE Syst. J.}, vol.~6, no.~4, pp. 616--626, Dec.
  2012.

\bibitem{kim}
T.~T. Kim and H.~V. Poor, ``Strategic protection against data injection attacks
  on power grids,'' \emph{IEEE Trans. Smart Grid}, vol.~2, no.~2, pp. 326--333,
  2011.

\bibitem{kosut}
O.~Kosut, L.~Jia, R.~J. Thomas, and L.~Tong, ``Malicious data attacks on the
  smart grid,'' \emph{IEEE Trans. Smart Grid}, vol.~2, no.~4, pp. 645--658,
  2011.

\bibitem{book}
A.~Abur and A.~Exp{\'o}sito, \emph{Power System State Estimation: Theory and
  Implementation}, ser. Power Engineering.\hskip 1em plus 0.5em minus
  0.4em\relax Marcel Dekker, 2004.

\bibitem{dist}
A.~Tajer, S.~Kar, H.~V. Poor, and S.~Cui, ``Distributed joint cyber attack
  detection and state recovery in smart grids,'' in \emph{Proc. 2nd IEEE Int.
  Conf. Smart Grid Communications}, Brussels, Belgium, Oct. 2011, pp. 202--207.

\bibitem{OEYKP_smc_12_2}
M.~Ozay, I.~Esnaola, F.~T.~Y. Vural, S.~R. Kulkarni, and H.~V. Poor,
  ``Distributed models for sparse attack construction and state vector
  estimation in the smart grid,'' in \emph{Proc. 3rd IEEE Int. Conf. Smart Grid
  Communications}, Tainan City, Taiwan, Nov. 2012.

\bibitem{fully}
L.~Xie, D.-H. Choi, S.~Kar, and H.~V. Poor, ``Fully distributed state
  estimation for wide-area monitoring systems,'' \emph{IEEE Trans. Smart Grid},
  vol.~3, no.~3, pp. 1154--1169, Sept. 2012.

\bibitem{dist2}
F.~Pasqualetti, R.~Carli, and F.~Bullo, ``A distributed method for state
  estimation and false data detection in power networks,'' in \emph{Proc. 2nd
  IEEE Int. Conf. Smart Grid Communications}, Brussels, Belgium, Oct. 2011, pp.
  469--474.

\bibitem{hierer}
Q.~Yang, J.~Yang, W.~Yu, N.~Zhang, and W.~Zhao, ``On a hierarchical false data
  injection attack on power system state estimation,'' in \emph{Proc. IEEE
  Global Communications Conference (GLOBECOM 2011)}, Houston, TX, USA, Dec.
  2011, pp. 1--5.

\bibitem{mitig}
O.~Vukovic, K.~C. Sou, G.~Dan, and H.~Sandberg, ``Network-aware mitigation of
  data integrity attacks on power system state estimation,'' \emph{IEEE J. Sel.
  Areas Commun.}, vol.~30, no.~6, pp. 1108--1118, July 2012.

\bibitem{review1}
Y.-F. Huang, S.~Werner, J.~Huang, N.~Kashyap, and V.~Gupta, ``State estimation
  in electric power grids: Meeting new challenges presented by the requirements
  of the future grid,'' \emph{IEEE Signal Process. Mag.}, vol.~29, no.~5, pp.
  33--43, Sept. 2012.

\bibitem{review2}
W.~Wang and Z.~Lu, ``Cyber security in the smart grid: Survey and challenges,''
  \emph{Comput. Netw.}, 2013, in press.

\bibitem{cn1}
D.~P. Chassin and C.~Posse, ``Evaluating north american electric grid
  reliability using the {B}arab{\'a}si-{A}lbert network model,'' \emph{Physica
  A: Statistical Mechanics and its Applications}, vol. 355, no. 2-4, pp.
  667--677, Sep. 2005.

\bibitem{admm}
S.~Boyd, N.~Parikh, E.~Chu, B.~Peleato, and J.~Eckstein, ``Distributed
  optimization and statistical learning via the alternating direction method of
  multipliers,'' \emph{Found. Trends Mach. Learn.}, vol.~3, no.~1, pp. 1--122,
  Jan. 2011.

\bibitem{sel}
E.~Ghadimi, A.~Teixeira, I.~Shames, and M.~Johansson, ``On the optimal
  step-size selection for the alternating direction method of multipliers,'' in
  \emph{Proc. 3rd IFAC Workshop on Distributed Estimation and Control in
  Networked Systems}, Santa Barbara, CA, USA, Sept. 2012.

\bibitem{rate}
B.~He and X.~Yuan, ``On the o(1/n) convergence rate of the {D}ouglas-{R}achford
  alternating direction method,'' \emph{SIAM J. Numer. Anal.}, vol.~50, no.~2,
  pp. 700--709, Apr. 2012.

\bibitem{candes}
E.~J. Cand{\`e}s and T.~Tao, ``Decoding by linear programming,'' \emph{IEEE
  Trans. Inf. Theory}, vol.~51, no.~12, pp. 4203--4215, 2005.

\bibitem{donoho}
D.~L. Donoho, ``Compressed sensing,'' \emph{IEEE Trans. Inf. Theory}, vol.~52,
  no.~4, pp. 1289--1306, 2006.

\bibitem{mp}
R.~D. Zimmerman, C.~E. Murillo-S\'{a}nchez, and R.~J. Thomas, ``{MATPOWER:
  Steady-state perations, planning, and analysis tools for power systems
  research and education},'' \emph{IEEE Trans. Power Syst.}, vol.~26, no.~1,
  pp. 12--19, Feb. 2011.

\bibitem{lasso}
R.~Tibshirani, ``Regression shrinkage and selection via the {LASSO},'' \emph{J.
  R. Stat. Soc. (Series B)}, vol.~58, pp. 267--288, 1996.

\bibitem{ridge}
Z.~Zhang, G.~Dai, C.~Xu, and M.~I. Jordan, ``Regularized discriminant analysis,
  ridge regression and beyond,'' \emph{J. Mach. Learn. Res.}, vol.~11, pp.
  2199--2228, Aug. 2010.

\bibitem{tikhonov}
A.~Tikhonov and V.~Arsenin, \emph{Solutions of {I}ll-posed {P}roblems}, ser.
  Scripta {S}eries in {M}athematics.\hskip 1em plus 0.5em minus 0.4em\relax
  Winston, 1977.

\bibitem{tik_power}
M.~de~Almeida, A.~Garcia, and E.~Asada, ``Regularized least squares power
  system state estimation,'' \emph{IEEE Trans. Power Syst.}, vol.~27, no.~1,
  pp. 290--297, Feb. 2012.

\bibitem{tau}
F.~Pasqualetti, R.~Carli, and F.~Bullo, ``A distributed method for state
  estimation and false data detection in power networks,'' in \emph{Proc. 2nd
  IEEE Int. Conf. Smart Grid Communications}, Brussels, Belgium, Oct. 2011, pp.
  469--474.

\bibitem{Sundeep}
S.~Rangan, A.~K. Fletcher, and V.~K. Goyal, ``Asymptotic analysis of {MAP}
  estimation via the replica method and applications to compressed sensing,''
  \emph{IEEE Trans Inf. Theory}, vol.~58, no.~3, pp. 1902--1923, 2012.

\bibitem{bach_lasso}
F.~R. Bach, ``Consistency of the group {LASSO} and multiple kernel learning,''
  \emph{J. Mach. Learn. Res.}, vol.~9, pp. 1179--1225, Jun. 2008.

\end{thebibliography}


\end{document}